\documentclass[PAPER, cernpreprint,atlasdraft=false, coverpage=false, UKenglish, texmf, orcidlogo]{atlasdoc}
\usepackage[]{hyperref}
\newcounter{chapter}

\newcommand{\chapter}[1]
{ {\centering
\addtocounter{chapter}{1} \Large \underline{\textbf{ \color{blue} Chapter \thechapter: ~#1}} }
\addcontentsline{toc}{section}{ \color{blue} Chapter:~\thechapter~~ #1}
}

\usepackage{atlaspackage}
\usepackage{atlasbiblatex}
\usepackage{atlascontribute}

\usepackage{atlasphysics}
\usepackage{comment}
\graphicspath{{logos/}{figures/}}

\usepackage{ANA-JETM-2024-03-PAPER-defs}
\AtlasTitle{The calibration of large-radius jets using the Run 2 dataset with the ATLAS detector}

\AtlasVersion{2.3}

\AtlasAbstract{%
The ATLAS experiment has developed an extensive procedure based on Monte Carlo simulation and collision data to calibrate the response of large-radius jets. This paper describes the Run~2 calibration of large-radius jets reconstructed from Unified Flow Objects and groomed with the Soft Drop algorithm, using data produced in proton--proton collisions at $\sqrt{s}=13$~TeV. Results are presented for the Monte Carlo-based correction and for the {\em in situ} methods that are used to validate the jet energy and mass scales and their resolutions in collision data. The residual uncertainty on the jet energy and mass scales is close to 1\% for transverse momenta up to 1~TeV and is 2--3\% up to 2~TeV, enabling sensitive searches to beyond the Standard Model particles and precise Standard Model measurements with boosted objects.
}

\AtlasRefCode{JETM-2024-03}
\AtlasJournal{Eur.Phys.J.C}

\AtlasAuthorContributor{ATLAS Collaboration}{}{}

\AtlasRefCode{JETM-2024-03}

\AtlasNote{JETM-2024-03}

\PreprintIdNumber{CERN-EP-2026-196}

\hypersetup{pdftitle={ATLAS document},pdfauthor={The ATLAS Collaboration}}

\begin{document}

\maketitle

\section{Introduction}
\label{sec:intro}
%

%

%
%

The study of massive particles such as top quarks and $W$, $Z$ and Higgs bosons produced in high-energy proton--proton ($pp$) collisions at the Large Hadron Collider (LHC) is a primary objective of the ATLAS experiment~\cite{GENR-2019-02}.
When such particles decay hadronically they produce showers of particles, which can be clustered into jets. In the following, jets built with the anti-$k_t$ jet clustering algorithm~\cite{Cacciari:2008gp} implemented in the \textsc{FastJet} package~\cite{Cacciari:2006sm,Fastjet} are considered.
Whenever the transverse momentum \pT of the decaying particle significantly exceeds its rest mass, the decay products are highly collimated and can no longer be resolved easily as multiple separate small-radius (\enquote{small-$R$}) jets, i.e. jets defined with a radius parameter of typically $R=0.4$. It is then advantageous to reconstruct a single jet with a large radius parameter of typically $R=1.0$ (a \enquote{large-$R$’} jet) that contains all decay products of the massive particle~\cite{Butterworth:2008iy}. Large-$R$ jets used to reconstruct such \enquote{boosted objects} can be distinguished (\enquote{tagged}) from the copious quark/gluon-initiated multijet background using jet substructure observables~\cite{JETM-2018-03,PERF-2017-04}. Large-$R$ jets and jet substructure are key tools in searches for massive new states and in the characterisation of the high-energy tail of Standard Model (SM) processes~\cite{Kogler:2018hem,Abdesselam:2010pt}.

The energy of reconstructed jets is calibrated to correct for the impact of detector inefficiencies, non-compensation of the calorimeter system, dead material and similar effects.  The calibration procedure is based on a combination of corrections derived from Monte Carlo (MC) simulation and data. The MC calibration matches reconstructed jets to the corresponding jet made from generated detector-stable particles and restores the scale to that of the particle-level jet. The {\em in situ} calibration, derived from collision data, validates and improves the agreement of data and MC\@. The correction factors from the MC calibration are applied to both data and MC events, while the {\em in situ} corrections are applied to data only to bring the data to the same energy scale as the MC particle-level. The calibration for large-$R$ jets follows the main steps of the chain applied to small-$R$ jets, extending it with an explicit MC and {\em in situ} calibration of the jet invariant mass~\cite{JETM-2018-02}.

In Ref.~\cite{JETM-2018-02}, the calibration of large-$R$ jet energy and mass using data collected during Run~2 of the LHC, from 2015 to 2018, was derived for a baseline jet collection reconstructed from topological clusters (\enquote{topoclusters})~\cite{PERF-2014-07} that was calibrated using local cell weighting (LCW) and groomed with the trimming algorithm~\cite{Krohn:2009th}. In this paper, the calibration for an updated baseline large-$R$ jet definition, reconstructed from Unified Flow Objects (UFOs)~\cite{JETM-2018-06}, is presented. The UFOs combine information from the calorimeter system with charged-particle tracks reconstructed in the inner detector, extending the particle-flow (PFlow) reconstruction~\cite{PERF-2015-09} with a track-aided Track-CaloCluster (TCC) splitting~\cite{HDBS-2018-31}.
To reduce the dependence of the jet radiation patterns on phenomena which are difficult to model precisely, such as soft radiation from the jet constituents, the underlying event and the effect of multiple $pp$ interactions in the same and neighbouring bunch crossings (\enquote{pile-up}),  a soft drop grooming~\cite{Larkoski:2014wba} is applied to the jet. Pile-up is further mitigated using the constituent subtraction~\cite{Berta:2019hnj} and the SoftKiller~\cite{Cacciari:2014gra} algorithms.

This document details the most refined calibration applied in analyses ATLAS data collected from 2015 to 2018 (Run~2) and forms the basis for the initial jet calibration used for data collected from 2022 to 2026 (Run~3). It builds on the experience gained through years of more refined and increasingly precise jet calibrations~\cite{PERF-2014-02, JETM-2018-05,JETM-2022-01}.
The paper is structured as follows. In Section~\ref{sec:atlasdet} the ATLAS detector is introduced, together with the data and MC simulations used. More details of the reconstruction of the large-$R$ jets used are provided in Section~\ref{sec:Jet_Reco}. The MC-based calibration is reported in Section~\ref{sec:mccalib}, while measurements entering the \emph{in situ} calibration are presented in Section~\ref{sec:insitujes} for the jet energy scale, in Section~\ref{sec:insitujer} for the jet energy resolution and in Section~\ref{sec:insitujmsr} for the jet mass.
Combinations of the jet response measurements are explained in Section~\ref{sec:comb_and_uncs}, where also additional uncertainties affecting the calibration procedure are discussed. The paper results are finally summarised in Section~\ref{sec:conclusions}.


\section{The ATLAS detector and data set}
\label{sec:atlasdet}

\subsection{The ATLAS detector}
The ATLAS experiment~\cite{PERF-2007-01} at the LHC is a multipurpose particle detector
with a forward--backward symmetric cylindrical geometry and a near \(4\pi\) coverage in
solid angle.\footnote{ATLAS uses a right-handed coordinate system with its origin at the nominal interaction point (IP)
in the centre of the detector and the \(z\)-axis along the beam pipe.
The \(x\)-axis points from the IP to the centre of the LHC ring,
and the \(y\)-axis points upwards.
Polar coordinates \((r,\phi)\) are used in the transverse plane,
\(\phi\) being the azimuthal angle around the \(z\)-axis.
The pseudorapidity is defined in terms of the polar angle \(\theta\) as \(\eta = -\ln \tan(\theta/2)\) and is equal to the rapidity
$ y = \frac{1}{2} \ln \left( \frac{E + p_z}{E - p_z} \right) $ in the relativistic limit.
Angular distance is measured in units of \(\Delta R \equiv \sqrt{(\Delta y)^{2} + (\Delta\phi)^{2}}\).}
It consists of an inner tracking detector surrounded by a thin superconducting solenoid
providing a \qty{2}{\tesla} axial magnetic field, electromagnetic and hadronic calorimeters, and a muon spectrometer.
The inner tracking detector covers the pseudorapidity range \(|\eta| < 2.5\).
It consists of silicon pixel, silicon microstrip, and transition radiation tracking detectors.
Lead/liquid-argon (LAr) sampling calorimeters provide electromagnetic (EM) energy measurements
with high granularity within the region \(|\eta|< 3.2\).
A steel/scintillator-tile hadronic calorimeter covers the central pseudorapidity range (\(|\eta| < 1.7\)).
The endcap and forward regions are instrumented with LAr calorimeters
for EM and hadronic energy measurements up to \(|\eta| = 4.9\).
The muon spectrometer surrounds the calorimeters and is based on
three large superconducting air-core toroidal magnets with eight coils each.
The field integral of the toroids ranges between \num{2.0} and \qty{6.0}{\tesla\metre}
across most of the detector.
The muon spectrometer includes a system of precision tracking chambers up to \(|\eta| = 2.7\) and fast detectors for triggering up to \(|\eta| = 2.4\).
The luminosity is measured mainly by the LUCID--2~\cite{LUCID2} detector which is located close to the beampipe.
A two-level trigger system was used to select events~\cite{TRIG-2016-01}.
The first-level trigger is implemented in hardware and used a subset of the detector information
to accept events at a rate close to \qty{100}{\kHz}.
This is followed by a software-based trigger that
reduced the accepted rate of complete events to \qty{1.25}{\kHz} on average
depending on the data-taking conditions.
A software suite~\cite{SOFT-2022-02} is used in data simulation, in the reconstruction
and analysis of real and simulated data, in detector operations, and in the trigger and data acquisition
systems of the experiment.

Data produced in $pp$ collisions at a centre-of-mass energy of $\sqrt{s} = 13~\TeV$ and collected in the Run~2 period of the LHC data-taking are used. Data must satisfy requirements based on beam and detector conditions and data quality~\cite{DAPR-2018-01}. The complete Run~2 data set corresponds to an integrated luminosity of $140.1 \pm 1.2~\text{fb}^{-1}$~\cite{DAPR-2021-01}.
The luminosity calibration relies on the LUCID-2 detector~\cite{LUCID2} for the primary luminosity measurements, complemented by measurements using the inner detector and calorimeters.

\subsection{Monte Carlo simulation}
\label{sec:MC}

Samples of simulated data are produced using MC event generators interfaced to a detailed description of the detector response~\cite{SOFT-2010-01} based on \GEANT~\cite{Agostinelli:2002hh}.  Additional $pp$ interactions are simulated with \PYTHIA[v8.186]~\cite{Sjostrand:2007gs} with the A3 tune~\cite{ATL-PHYS-PUB-2016-017} and overlaid onto hard-scattering processes. Simulated events are weighted based on the number of pile-up events in the MC to match the observed distribution of the average number of interactions per bunch crossing. The same reconstruction algorithms are applied to data and MC events.
Several different MC generators are employed to simulate the different physics processes that are used throughout the calibration chain.

The majority of high-\pT jets at the LHC stem from quantum chromo dynamic (QCD) multijet production initiated by light quarks and gluons. This process is used to derive the MC calibration and plays an important role in several steps of the \emph{in situ} calibration chain.
Multijet production is simulated using a variety of MC simulations.
One sample is simulated using \PYTHIA[v8.230]~\cite{Sjostrand:2014zea} with leading-order (LO) matrix elements (ME) for dijet that are matched to the parton shower. The \NNPDF[2.3lo] set~\cite{Ball:2012cx} of parton distribution functions (PDF) and the A14 set of tuned parameters~\cite{ATL-PHYS-PUB-2014-021} are used in the ME generation, the parton shower, and the simulation of the multi-parton interactions.
Multijet production at next-to-leading-order (NLO) accuracy is produced with \POWHEGBOX[v2]~\cite{Nason:2004rx,Frixione:2007vw,Alioli:2010xa,Alioli:2010xd} interfaced to \PYTHIA[8].
The \NNPDF[3.0nlo]~\cite{Ball:2014uwa} PDF is used, together with \PYTHIA with the A14 tune and the
\NNPDF[2.3lo] PDF for the shower and multi-parton interactions.
Multijet production is also simulated at NLO with \HERWIG[7.1.3]~\cite{Bellm:2017jjp}, using the
\MMHT[nlo]~\cite{Harland-Lang:2014zoa} PDF set for the ME calculation and the cluster model~\cite{Winter:2003tt} for the hadronisation algorithm.
Two sets of \HERWIG samples are generated, where one makes use of the default parton shower with angular ordering and the other uses the dipole shower as an alternative, hence enabling the study of the effects of using different parton shower models.
The \HERWIG angular-ordered shower algorithm is also used interfaced to \POWHEG, similarly to the \POWPY simulation already introduced.
Two multijet production samples simulated using the \SHERPA[2.2.5]~\cite{Bothmann:2019yzt} generator are used to study the effect of the hadronisation model choice.  The matrix element calculation was included for the \(2\rightarrow2\) process at leading order, and the default \SHERPA parton
shower~\cite{Schumann:2007mg} based on Catani--Seymour dipole factorisation was used for the showering with \pT
ordering, using the \CT[14nnlo] PDF set~\cite{Dulat:2015mca}.
The first of these samples makes use of the dedicated \SHERPA AHADIC model for hadronisation~\cite{Winter:2003tt} while the second sample is simulated using the \SHERPA interface to
the Lund string fragmentation model of \PYTHIA[6]~\cite{Sjostrand:2006za} and its decay tables.

Associated production of a heavy vector boson ($V=W,Z$) or a photon with jets is important for the \emph{in situ} calibration. Production of a $V$ boson with jets is simulated using the \POWHEGBOX~\cite{Alioli:2008gx} ME generator at NLO QCD accuracy, interfaced with \PYTHIA[v8.186]. %
An alternative sample is created with \SHERPA~2.2.1~\cite{Gleisberg:2008ta} with NLO QCD accuracy in the ME calculation for up to two partons, and LO ME
for up to four partons.
The nominal MC generator for production of a photon and jets is \PYTHIA[v8.230] at LO in QCD, while an alternative sample is simulated with \SHERPA~2.2.2 at NLO in QCD. The $Z$+jets samples are produced with the \NNPDF[3.0NNLO] PDF set, while the $\gamma$+jets ones are produced with \NNPDF[2.3LO].
The production of diboson ($WW/WZ/ZZ$) events in association with jets is simulated using a similar setup as for the $V$+jets samples,
with a dedicated tune from the \SHERPA authors~\cite{Bothmann:2019yzt}.

The production of top-quark pairs (\ttbar) and top-quark in asosciation with a $W$ boson ($tW$) is modelled using the \POWHEGBOX[v2] generator at NLO~\cite{Frixione:2007nw} with the \NNPDF[3.0nlo] PDF set, setting the value of the top-quark mass $m_\text{top}=172.5~\GeV$. The events are interfaced
to \PYTHIA[v8.230] to model the parton shower,
hadronisation, and underlying event, with parameters set according to the A14 tune and using the \NNPDF[2.3lo] set of PDFs. The decays of bottom and charm hadrons are performed by \EVTGEN[1.6.0]~\cite{Lange:2001uf}.
Alternative MC $t\bar t$ predictions are used to estimate the uncertainties in the modelling of the initial- and final-state radiation, of the ME calculation and of the showering and hadronisation. Alternative \ttbar \POWPY[8] predictions are employed with the renormalisation and factorisation scales varied by a factor of 2, using the low radiation variation of the A14 tune, corresponding to reduced parton-shower radiation.
The uncertainty on the choice of the \hdamp parameter\footnote{The
\hdamp parameter is a resummation damping factor and one of the
parameters that controls the matching of \POWHEG ME to  the parton shower and thus effectively regulates the high-\pT radiation against which the \ttbar system recoils.} is estimated comparing the nominal  \ttbar \POWPY[8] MC simulation generated with $\hdamp=1.5\times m_\text{top}$ to an alternative sample with $\hdamp=3 \times m_\text{top}$.
The uncertainty in the simulation of the parton shower is evaluated by employing a sample interfaced with \HERWIG[7.04]~\cite{Bahr:2008pv,Bellm:2015jjp},
using the H7UE set of tuned parameters~\cite{Bellm:2015jjp} and the
\MMHT[lo] PDF set~\cite{Harland-Lang:2014zoa}.

The topology dependence of the simulation-based calibration is studied on beyond the SM (BSM) processes, that produce jets originating from $W/Z$ bosons and top quarks decays with $p_\text{T}$ up to high values that are only scarcely populated by SM processes.
Jets from $W$ bosons are produced using a BSM spin-1 $W'\to WZ \to qqqq$ model including only hadronic $W$ and $Z$ decays. The mass of the $W'$ boson ranges from 400~\gev{} to 5~\tev{}. The top quark jets are from a BSM $Z'\to t\bar{t}$ model where the top quark may decay either hadronically or semileptonically. The mass of the $Z'$ boson ranges from 400~\gev{} to 3~\tev. For both BSM models, the production is simulated using \PYTHIA[v8.235] with the A14 tune and the \NNPDF[2.3lo] PDF set.
Two $W(\rightarrow qq)\gamma$ MC samples are used to evaluate uncertainties related to the topology of jets initiated by a $W$ boson: the first sample is generated with \SHERPA[2.2.11] with ME at NLO QCD accuracy for up to one additional parton and LO accuracy for up to three additional parton emissions, while the second sample is generated with \MGNLO[2.2.2]~\cite{Alwall:2014hca},
using LO-accurate ME with up to four final-state partons, interfaced to \PYTHIA[8.186]. Both $W(\rightarrow qq)\gamma$ samples use the \NNPDF[3.0] PDF set, at NNLO and NLO respectively.

%


\section{Large-$R$ jet reconstruction}
\label{sec:Jet_Reco}


The large-$R$ jets considered in this paper are reconstructed using the anti-$k_t$ algorithm with a radius parameter $R=1.0$. %
Three large-$R$ jet definitions are used in this paper: UFO jets, calorimeter jets and particle-level (or \enquote{truth}) jets.

The large-R UFO jets combine two types of signals: charged-particle-tracks from the inner detector %
and topoclusters.
The UFO reconstruction algorithm begins by applying the ATLAS PFlow algorithm on tracks and topoclusters to define a set of PFlow objects (PFOs), which are considered for jet clustering. To reduce the effect of pile-up, charged PFOs matched to pile-up vertices are rejected and the constituent subtraction and SoftKiller algorithms are applied to the set of PFOs.
A modified version of the TCC splitting algorithm is then applied to the remaining PFOs, where only tracks from the hard-scatter vertex are used as input
to the algorithm, in order to avoid pile-up instabilities, and any tracks that have been used to correct topocluster momenta in the PFlow algorithm are not considered.
This approach provides the maximum benefit from both the PFO and TCC algorithms. For a detailed description of the UFO algorithm, the reader is referred to Ref.~\cite{JETM-2018-06}.

To further reduce effects due to soft emissions, the underlying event and residual pile-up contamination on jet substructure calculations,
the large-$R$ UFO jet constituents are groomed with the soft drop algorithm. In this algorithm,
the constituents of the large-$R$ jet are reclustered using the C/A algorithm~\cite{Dokshitzer:1997in}, creating an angle-ordered jet clustering history. Then, the clustering sequence is traversed in reverse (starting from the widest-angled radiation and iterating towards the jet core), removing the softer branch until the splitting satisfies
\[
z \;>\; z_{\text{cut}}\!\left(\frac{\Delta R}{R}\right)^{\beta} \qquad \text{and}
\qquad
z = \frac{\min(p_{\mathrm{T1}},p_{\mathrm{T2}})}{p_{\mathrm{T1}}+p_{\mathrm{T2}}}\quad ,
\]
where $\Delta R$ is the angular separation of the two branches, each with momentum $p_\text{Ti}$, $R$ is the jet radius and values of $\beta = 1$ and $z_{\text{cut}} = 0.1$ were taken following the optimisation studies in Ref.~\cite{JETM-2018-06}.

Large-$R$ calorimeter jets were previously used in ATLAS~\cite{JETM-2018-02} and are reconstructed with
topoclusters calibrated with the LCW algorithm. %
The trimming algorithm %
is used for removing pile-up contamination and grooming the jet: the $R=1.0$ jet constituents are reclustered with the $k_t$ algorithm into sub-jets of radius $R_{\mathrm{sub}}=0.2$; then any sub-jet that carries less than 5\% of the original $p_{\mathrm{T}}$ is discarded while the retained sub-jets are assembled into the $R=1.0$ trimmed jet.

Particle-level large-$R$ jets are created by clustering stable particles in the simulation, which have a lifetime $\tau$ in the laboratory frame such that $c\tau>10$~mm. Muons and neutrinos are not considered as constituents for particle-level jets, as they do not leave significant energy deposition in the calorimeter.
For each reconstructed large-$R$ jet definition considered in this paper, the corresponding particle-level reference jet is built using the same grooming algorithm. Therefore, UFO soft drop jets are compared to soft drop particle-level jets, while LCW trimmed topocluster jets are compared to trimmed particle-level jets. This ensures that comparisons between jet definitions reflect differences between the reconstruction and calibration performance rather than a mismatch in the particle-level reference definition.


%
\section{Monte Carlo calibration}
\label{sec:mccalib}
%
%

%

%
Due to various inefficiencies and limitations inherent to the detectors, the energy, mass, and direction of the reconstructed jets may not align with those of the particle-level jets.
Accurate corrections are necessary before the jets can be utilised in physics analyses.
The first correction is obtained from simulation and is called the \enquote{MC calibration}.
Large-$R$ particle-level jets that use the same grooming technique as the large-$R$ jets reconstructed with the ATLAS detector are employed as a reference in this calibration.

The reconstructed large-$R$ jets in both data and MC are calibrated by applying corrections derived from simulated collision events to restore their four-momentum to that of the jets at the particle level. These corrections are applied first to the energy and pseudorapidity of the jet, and then to the jet mass following the numerical inversion procedure used also in previous ATLAS jet calibrations~\cite{JETM-2018-02, PERF-2014-02}.

Reconstructed large-$R$ jets with $\pt>100~\GeV$ are matched to particle-level large-$R$ jets with  with $\pt^\text{true}>100~\GeV$  using the angular distance
$\Delta R = \sqrt{(\Delta\eta)^2 + (\Delta\phi)^2}$~\cite{JETM-2018-05}.
To suppress ambiguities from nearby large-$R$ jets, isolation requirements are imposed on both truth and reconstructed jets:
a truth large-$R$ jet is required to have no additional truth large-$R$ jet within $\Delta R < 2.5$, and a reconstructed large-$R$ jet is required to have
no additional reconstructed large-$R$ jet within $\Delta R < 1.5$.
After applying these isolation criteria, a reconstructed jet is associated with a truth jet if their separation satisfies
$\Delta R(\text{true},\text{reco}) < 0.3$.
The individual jet energy response is defined as
$E^{\mathrm{reco}}/E^{\mathrm{true}}$, where $E^{\mathrm{reco}}$ is the energy calculated from jets after the full reconstruction procedure, and $E^{\mathrm{true}}$
is the energy of the particle-level jets.
The most probable value of the individual jet energy response distribution is estimated by fitting the core of the distribution with a gaussian function and taking the fitted peak position. %
The inverse of this quantity is referred to as the jet energy scale, JES\@. The jet mass scale, JMS, is defined analogously by applying the same gaussian-core fit procedure to the jet mass response distribution, ($m^{\mathrm{reco}}/m^{\mathrm{true}}$), and taking the fitted peak position.

\subsection{Monte Carlo jet energy calibration}

In order to calibrate the overall momenta of reconstructed jets, an energy correction factor is applied to the jet four-momentum as a multiplicative scale factor. This correction factor is determined as a function of the energy of the jet and the detector pseudorapidity $\eta_\text{det}$~\cite{JETM-2018-05}. The MC jet energy calibration is derived from \PYTHIA[8] multijet samples, dominated by large-$R$ UFO jets originating from fragmentation of light quarks and gluons. The calibration is derived for large-$R$ UFO jets for post-correction values of $\pt>150~\gev$ and $m^\text{reco} > 40~\gev$.

\begin{figure}[!ht]
\subfloat[]{\includegraphics[page=1,width=0.49\linewidth]{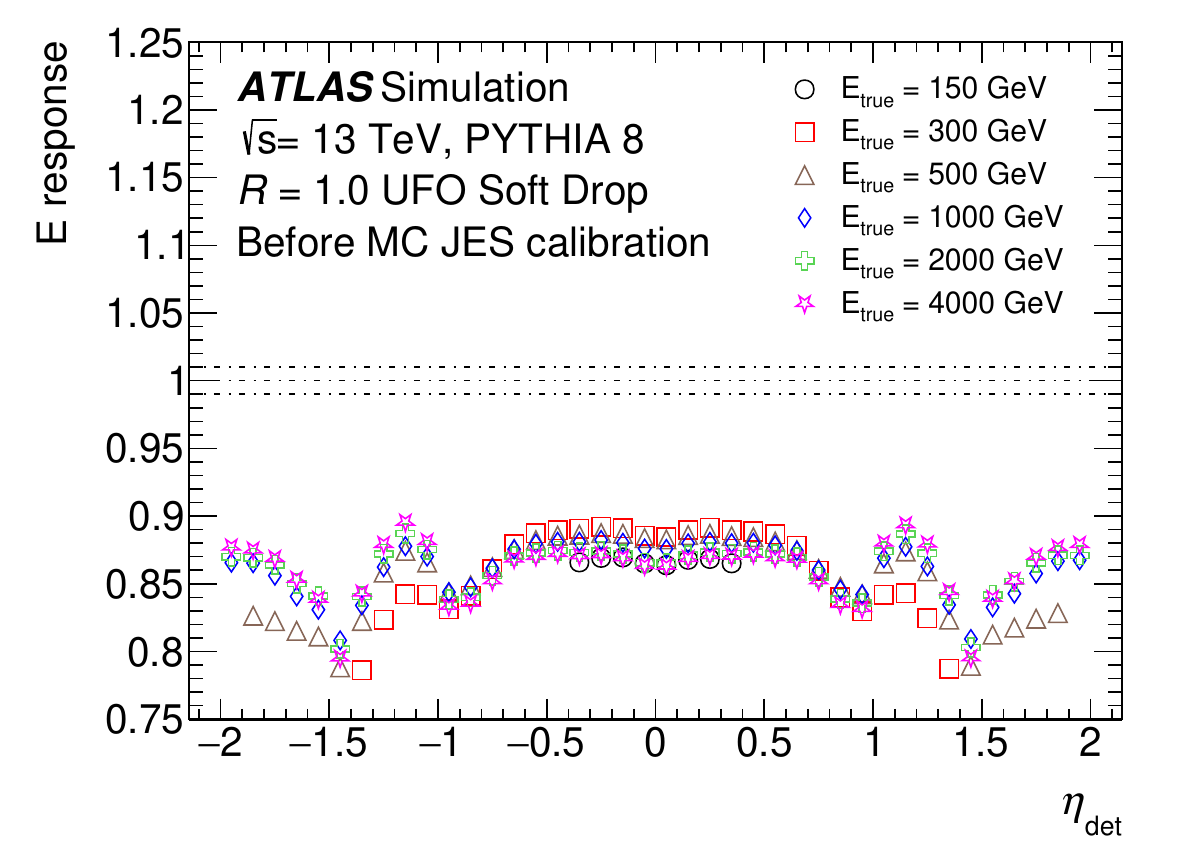}}
\subfloat[]{ \includegraphics[page=2, width=0.49\linewidth]{fig_01b.pdf}}
\caption{Large-$R$ UFO jet energy response (a) before and (b) after the Monte Carlo calibration is applied, as a function of $\eta_\text{det}$ for different true energies. The responses are evaluated in multijet events generated with \PYTHIA[8]. All data points include statistical uncertainties obtained from the gaussian fits, but they are too small to be visible.}
\label{fig:JES_Calib_uncalib}
\end{figure}

Figure~\ref{fig:JES_Calib_uncalib} shows the energy response for different detector pseudorapidities, $\eta_\text{det}$, and different jet energies, before and after the MC calibration. The energy response before calibration differs from unity by 10 -- 20\%, depending on jet energy and pseudorapidity.
The response after the MC calibration shows excellent closure. The energy scale after calibration is contained well within the horizontal dashed lines that represent a $1\%$ deviation from unity across the pseudorapidity range.

\subsection{Monte Carlo jet mass calibration}

One important goal for large-$R$ jets and the use of sophisticated grooming techniques is the accurate and precise reconstruction of the jet mass. To achieve this, an additional MC-based calibration step is applied to correct the reconstructed jet mass as a function of the energy of the jet and $|\eta_{\mathrm{det}}|$~\cite{JETM-2018-02,PERF-2012-02}.

Calibration factors for the reconstructed mass are derived both from multijet events and from boosted $W/Z$-boson events from the $W'\rightarrow WZ$ sample generated with \PYTHIA[8]. The use of large-$R$ jets originating from a $W/Z$ boson ($W/Z$ jets) provides a topology closer to the main physics use case of large-$R$ jets, namely the reconstruction of boosted massive particles, while multijet events provide the statistical precision and kinematic coverage at higher transverse
momenta. To achieve the best closure of the large-$R$ UFO jet mass scale, the calibration factors derived from $W/Z$-boson events are applied for $p_{\mathrm{T}}^{\mathrm{true}}<500$~\GeV, and those derived from
multijet events are applied for the remaining $p_{\mathrm{T}}$ range. This strategy is referred to as the \enquote{combined calibration}.

\begin{figure}[!ht]
\subfloat[]{\includegraphics[width=0.49\linewidth]{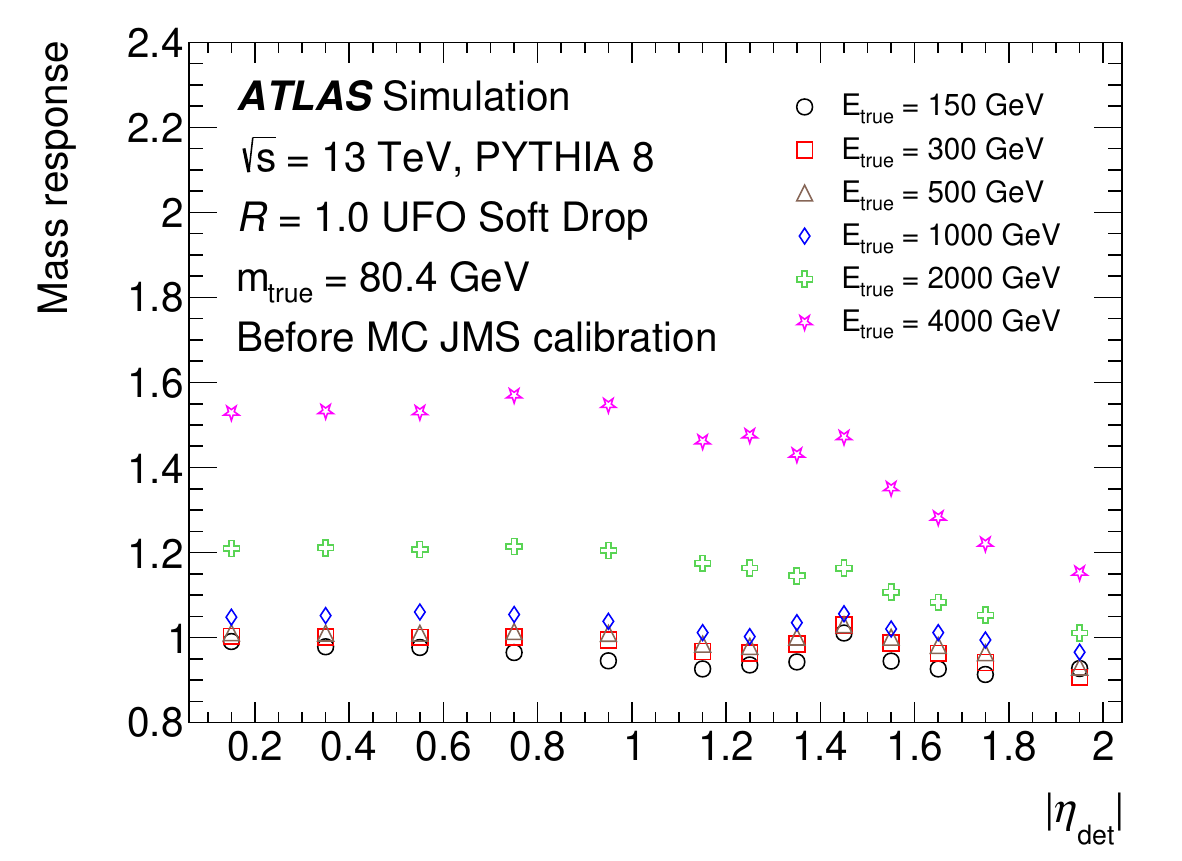}}
\subfloat[]{\includegraphics[width=0.49\linewidth]{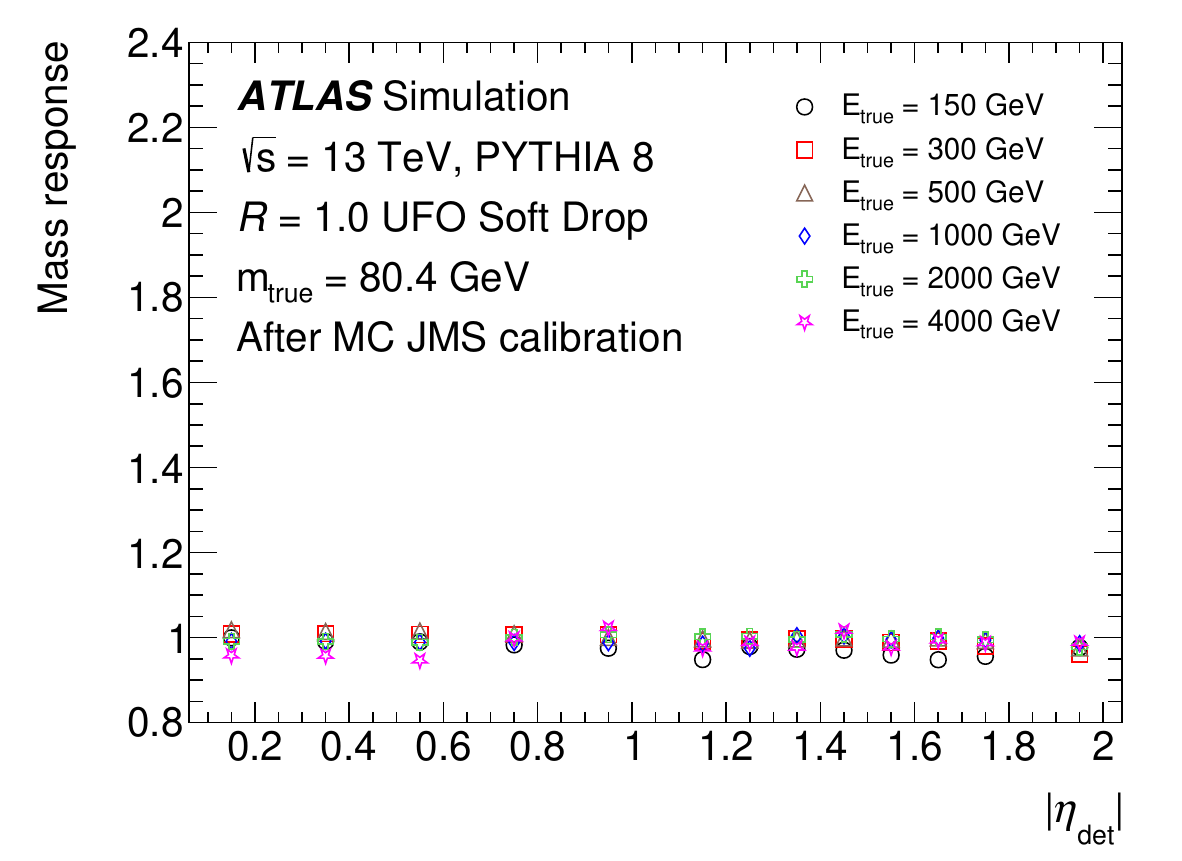}}\\
\subfloat[]{\includegraphics[width=0.49\linewidth]{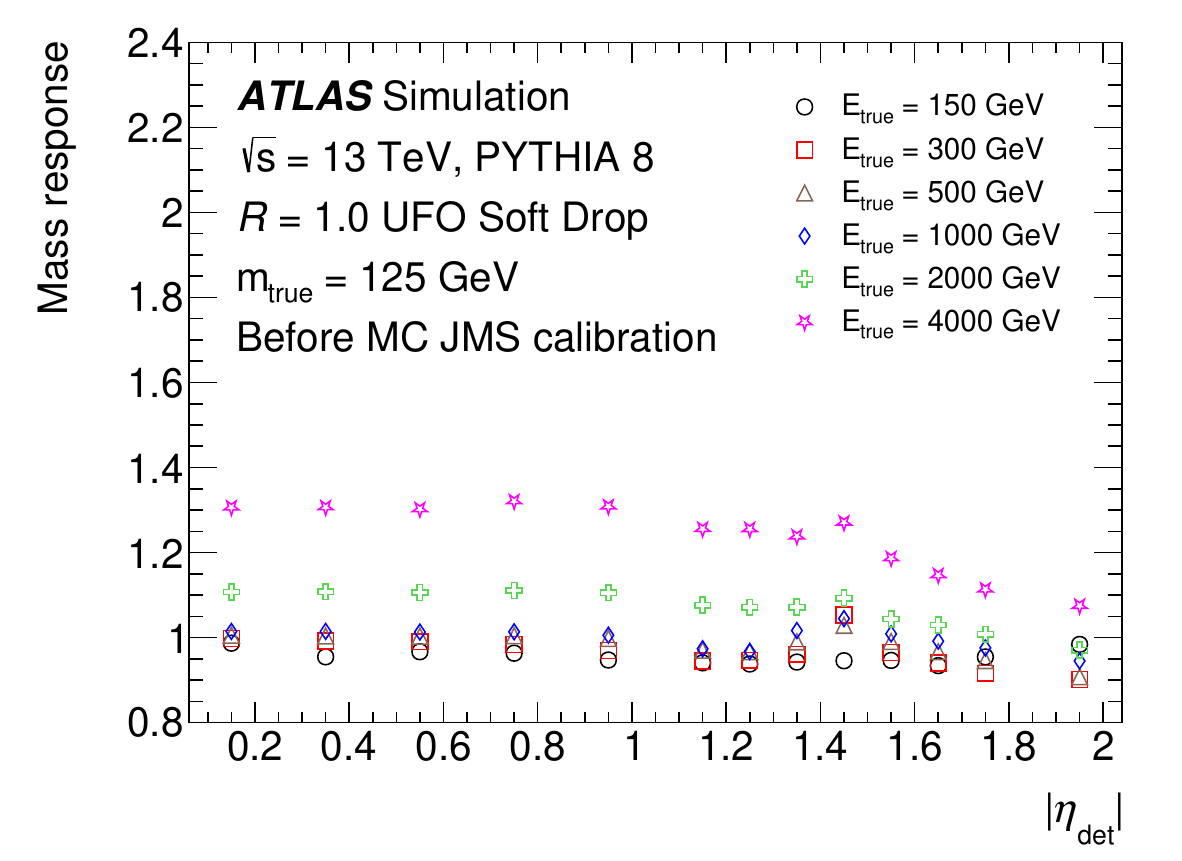}}
\subfloat[]{ \includegraphics[width=0.49\linewidth]{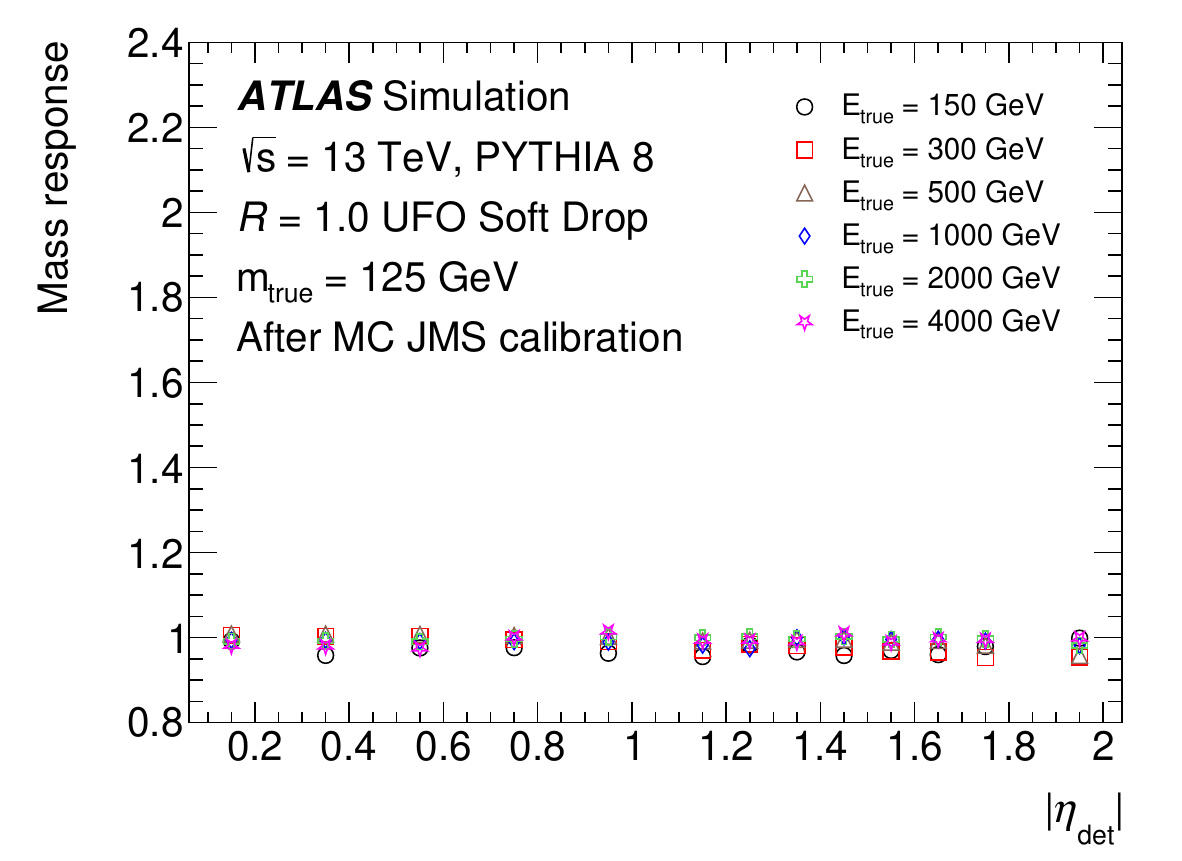}}\\
\subfloat[]{ \includegraphics[width=0.49\linewidth]{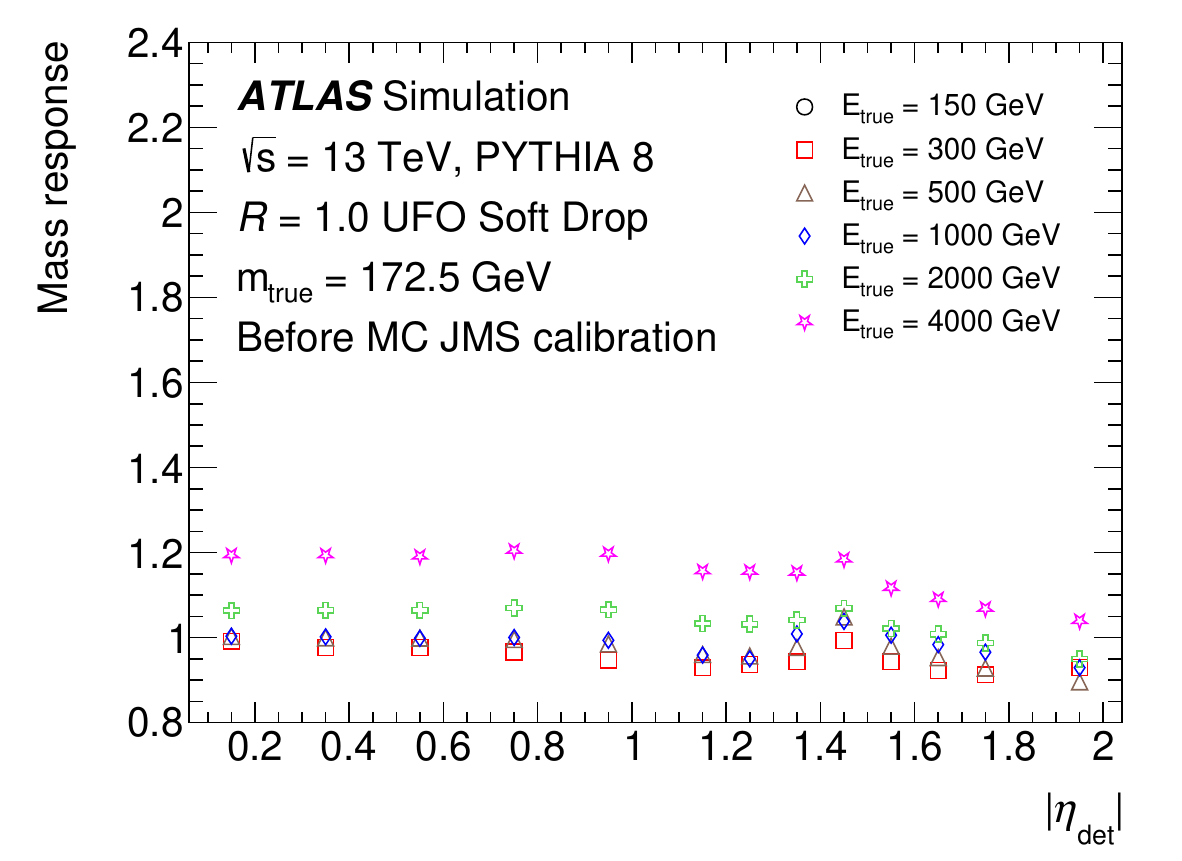}}
\subfloat[]{ \includegraphics[width=0.49\linewidth]{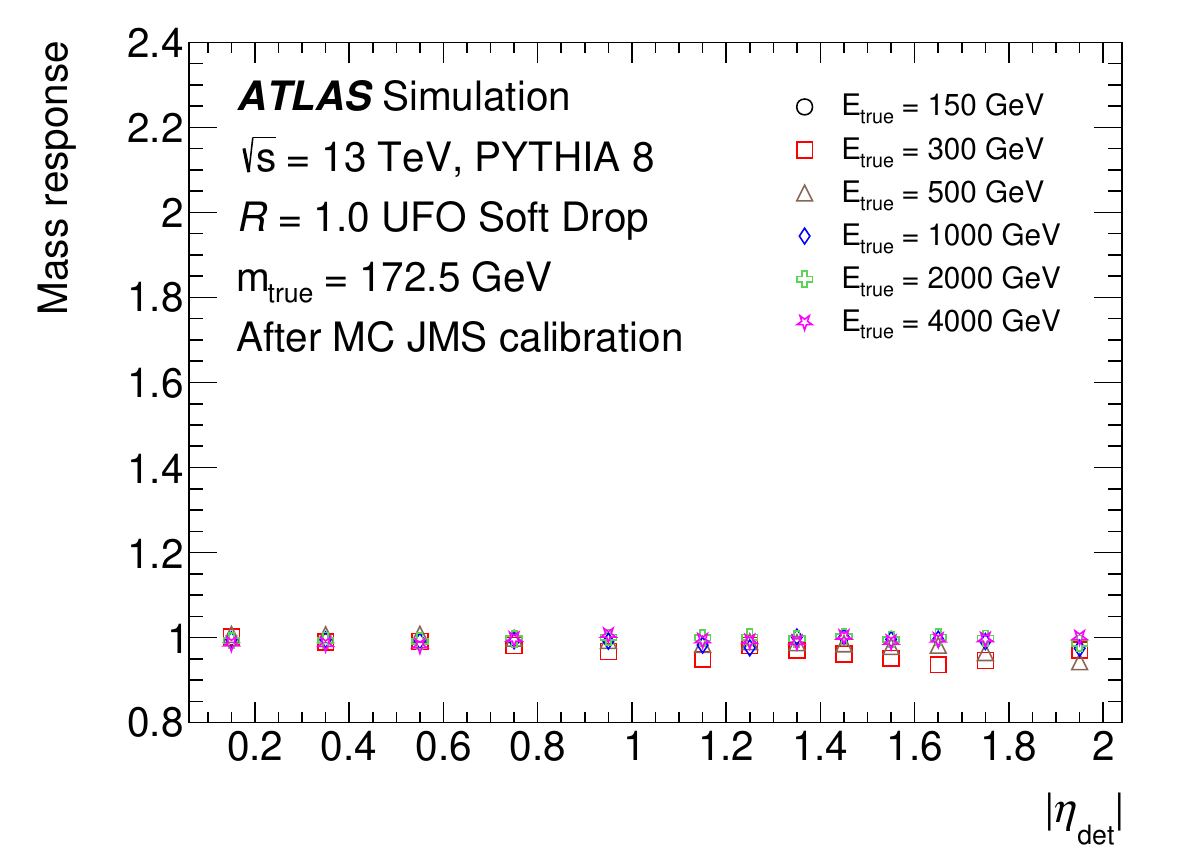}}
\caption{Large-$R$ UFO jet mass response  (a, c, e) before  and (b, d, f) after applying the combined Monte Carlo JMS calibration as a function of the detector pseudorapidity $|\eta_{\mathrm{det}}|$.
Three representative truth jet mass points of the response parameterisation are shown: (a, b) $m^{\mathrm{true}}=80.4$~\GeV, (c, d) 125~\GeV and (e, f) 172.5~\GeV; coloured markers correspond to different truth jet energies.
Figures are obtained from a joint combination of the two samples used to derive the calibration: hadronic $W/Z$ jets for $p_\text{T}^{\mathrm{true}}<500$~\GeV and QCD jets for $p_\text{T}^{\mathrm{true}}>500$~\GeV. The statistical uncertainties from the Gaussian-core fits are included but are too small to be visible.
}
\label{fig:JMS_Calib_uncalib}
\end{figure}

Figure~\ref{fig:JMS_Calib_uncalib} shows the large-$R$ UFO jet mass scale before and after the calibration as a function of $|\eta_{\mathrm{det}}|$ for several values of the truth jet energy.
Three truth jet masses are therefore considered: $80.4$~\GeV, $125$~\GeV, $172.5$~\GeV;
representing the approximate masses of the $W$ boson, Higgs boson, and top quark.
After the large-$R$ UFO jet mass calibration, good closure is achieved across the pseudorapidity range.
There is a residual non-closure that can reach up to 5\% for large-$R$ UFO jets with relatively low energy, even when the response is evaluated on the same samples used to derive the calibration. This is primarily due to imperfect fits arising from non-gaussian responses and threshold effects.

\begin{figure}[!ht]
\subfloat[]{\includegraphics[page=1, width=0.49\linewidth]{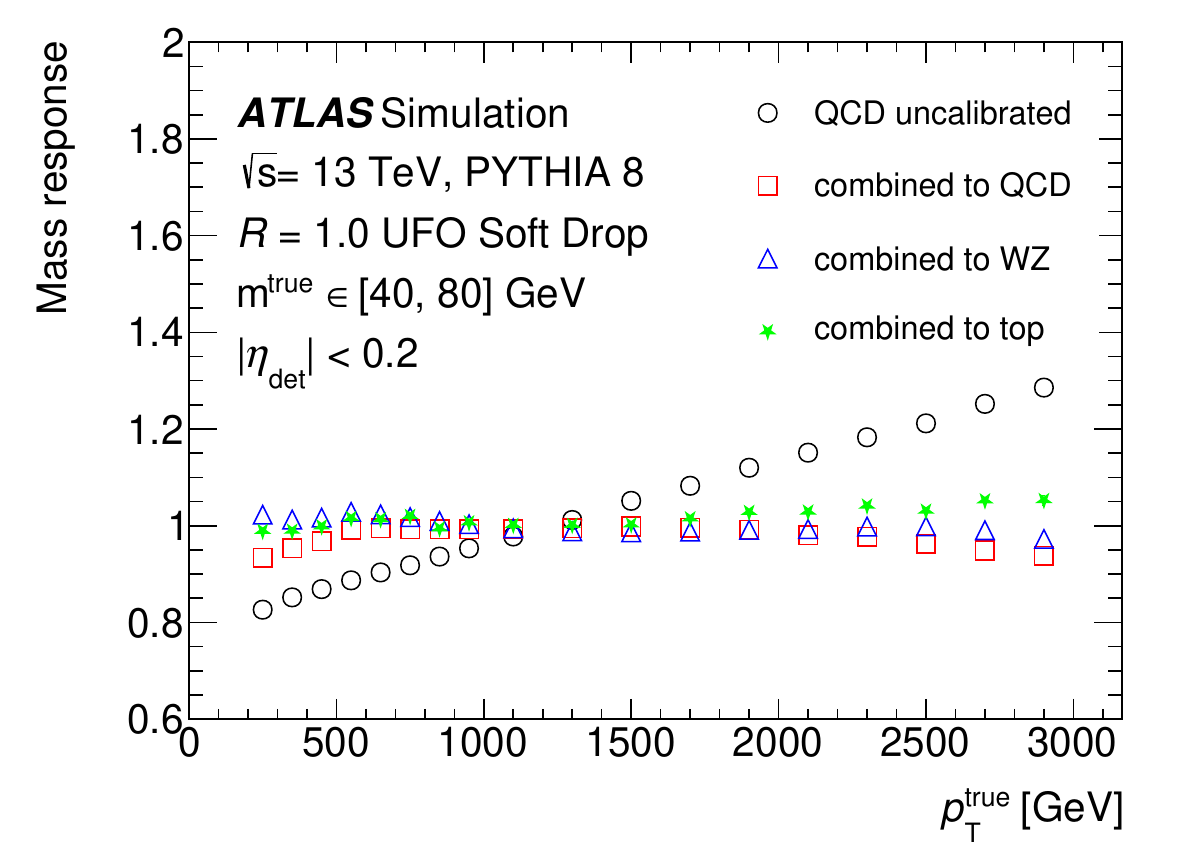}}
\subfloat[]{\includegraphics[page=2, width=0.49\linewidth]{fig_03b.pdf}}\\
\subfloat[]{\includegraphics[page=3, width=0.49\linewidth]{fig_03c.pdf}}
\subfloat[]{\includegraphics[page=4, width=0.49\linewidth]{fig_03d.pdf}}
\caption{Jet mass response as a function of $p_{\mathrm{T}}^{\mathrm{true}}$ for central jets ($|\eta_\text{det}| <0.2$) jets from several event topologies.
Four truth jet mass $m^{\mathrm{true}}$ intervals are shown: (a) $m^{\mathrm{true}}\in [40,80]$~\GeV, (b) $m^{\mathrm{true}}\in [80,100]$~\GeV, (c) $m^{\mathrm{true}}\in [100,140]$~\GeV and (d) $m^{\mathrm{true}}\in [140,200]$~\GeV.
The uncalibrated response of jets from a multijet simulation (QCD uncalibrated) is compared to the response obtained after applying the combined Monte Carlo jet mass calibration on jets generated in three processes: QCD multijet production (\enquote{combined to QCD}), production of hadronically-decaying $W/Z$ bosons from  $W' \to WZ$  (\enquote{combined to WZ}), and production of boosted top quarks from the $Z' \to t\bar{t}$ (\enquote{combined to top}). The statistical uncertainties from the Gaussian-core fits are included but are too small to be visible.}
\label{fig:JMS_Calib}
\end{figure}

Jets from hadronically-decaying massive particles have different shower topologies from the
multijet events used to derive the calibration factors for $p_{\mathrm{T}}^{\mathrm{true}} > 500~\GeV$.
The topology dependence of the mass response is therefore studied using simulated boosted
$W/Z$-boson and top-quark samples.
Figure~\ref{fig:JMS_Calib} presents the large-$R$ UFO jet mass scale after applying the combined calibration to simulated large-$R$ UFO jets in multijet (referred to as QCD), $W'\rightarrow W/Z$ and $Z'\rightarrow \ttbar$ samples. The large-$R$ UFO jet mass scale of uncalibrated jets from the multijet MC is also shown as a reference.
The jet mass scale is shown as a function of $p_{\mathrm{T}}^{\mathrm{true}}$ in four different $m^{\mathrm{true}}$ intervals: $[40,80]$~\GeV, $[80,100]$~\GeV, $[100,140]$~\GeV and $[140,200]$~\GeV, representing typical values for regions dominated respectively by jets from quarks or gluons, jets from $W/Z/H$ bosons and jets from top quarks.
Plots are shown for central jets ($|\eta_\text{det}|<0.2$) as a representative example but qualitatively similar results are obtained for other $|\eta_\text{det}|$ ranges.

\begin{figure}[!ht]
\subfloat[]{\includegraphics[page=1, width=0.49\linewidth]{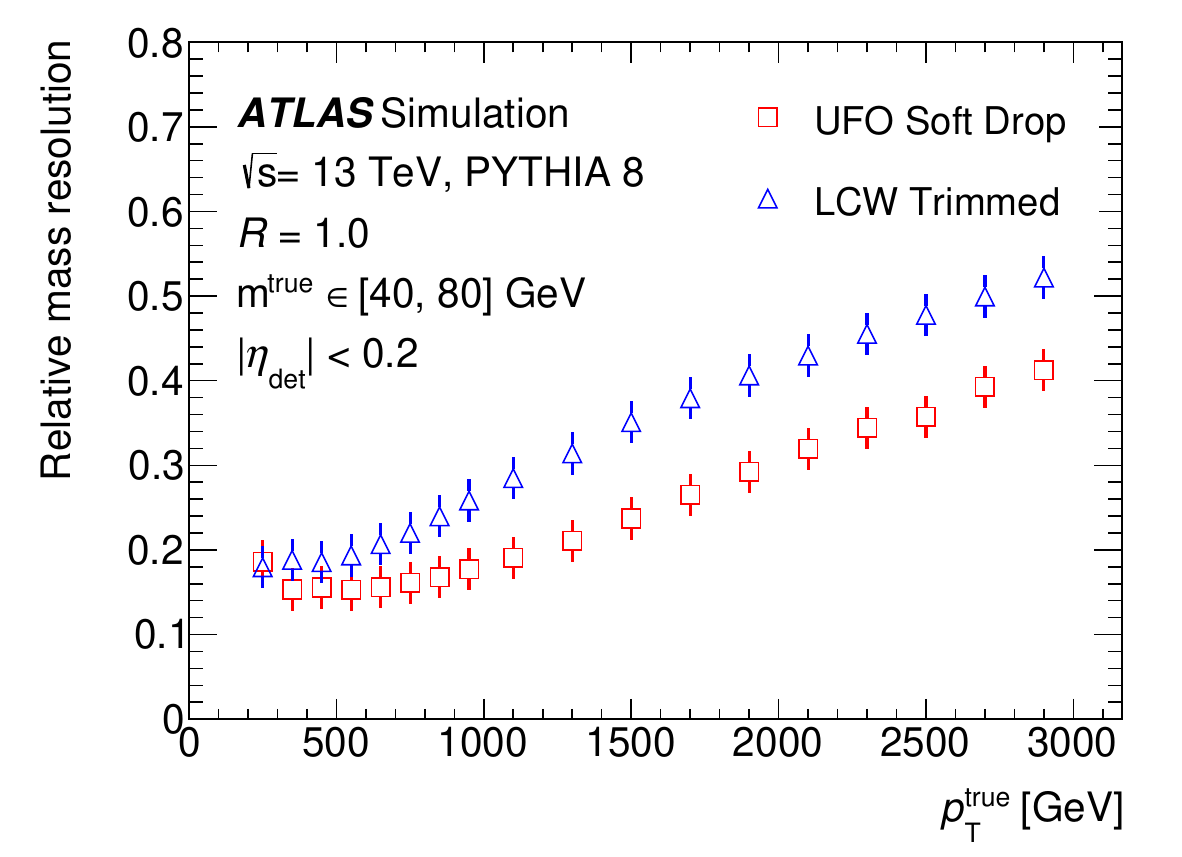}}
\subfloat[]{\includegraphics[page=3, width=0.49\linewidth]{fig_04b.pdf}}\\
\caption{Jet mass resolution as a function of $p_{\mathrm{T}}^{\mathrm{true}}$ for central jets ($|\eta_\text{det}|\in[0.0,0.2]$) in two intervals of the truth jet mass $m_{\mathrm{true}}$ (as indicated in each panel). Plots are shown for central jets as a representative example but the same checks were performed in the other $|\eta_\text{det}|$ bins, with no qualitatively different trends observed.
The resolution of large-$R$ UFO jets groomed with soft drop is shown, after applying the combined Monte Carlo jet-mass calibration. For comparison, the resolution obtained with the previous ATLAS large-$R$ jet definition based on calorimeter topological clusters and groomed with the trimming algorithm (``LCW Trimmed'') is also shown. The error bars quantify the uncertainty on that estimated width due to finite sample size and histogram binning.}
\label{fig:JMR_Calib}
\end{figure}

The large-$R$ jet mass resolution is evaluated from the width of the jet mass response distribution around the mode. In particular, the 68\% interquantile range, $\mathrm{IQR}{68}$, is used instead of a gaussian width or standard deviation to reduce the sensitivity to non-gaussian tails and asymmetric response shapes. The $\mathrm{IQR}{68}$ is divided by twice the mode of the response distribution to express the result as a relative resolution.
Figure~\ref{fig:JMR_Calib} illustrates the large-$R$ UFO jet mass resolution evaluated in multijet events after the combined mass calibration is applied.

The relative jet mass resolution is shown as a function of $p_{\mathrm{T}}^{\mathrm{true}}$ for two ranges of the true jet mass $m^{\mathrm{true}}$: $[40,80]~\GeV$ and $[100,140]~\GeV$. Results obtained for large-$R$ calorimeter jets are shown for comparison and labelled as \enquote{LCW trimmed} jets. The UFO jets with soft drop grooming show a significantly improved jet mass resolution in comparison to the trimmed calorimeter jets.

\FloatBarrier


%
\section{\emph{In situ} jet energy scale calibration}
\label{sec:insitujes}
%
%


\label{sec:jes_insitu}

The MC calibration relies on the correct description of the detector response in the simulation. The \emph{in situ} calibration validates such assumption and corrects the MC-calibrated large-$R$ UFO jets in data for eventual differences between data and MC, hence bringing the data to agree with the MC\@.
In this section, the methods used to derive the \emph{in situ} calibration for the energy (or transverse momentum) response are presented.
These methods use \pt conservation in events where a large-$R$ UFO jet recoils against a well-measured reference object.
The jet response is defined as the average ratio of the jet \pt to the reference object \pt in bins of the reference object \pt.
The response is transformed via numerical inversion from a function of reference object \pt to a function of the large-$R$ jet \pt~\cite{JETM-2018-02}.
The data-to-simulation jet response double ratio is then used to validate and correct the detector response for high-\pt large-$R$ UFO jets.

The reference object can be a photon, a di-muon or di-electron system from $Z$ boson decay, an ensemble of small-$R$ jets or a large-$R$ UFO jet measured in a region where the detector response to the jet energy is known to be very well modelled, such as the central region of the detector.%

Photons are defined from three-dimensional topological clusters of energy deposits in the electromagnetic calorimeter~\cite{EGAM-2018-01}, satisfying the ‘tight’ identification and isolation requirements~\cite{PERF-2017-02}.
Photons are required to have \(\pt > 25~\gev\) and to fall within \(|\eta| < 1.37 \) or \(1.52 < |\eta| < 2.37\),
avoiding the transition region between the barrel and endcap calorimeters.

Electron candidates are identified by matching energy deposits in the electromagnetic calorimeter with a corresponding track in the inner tracking detector.
They are required to satisfy ‘medium’ likelihood identification quality and ‘loose’ isolation requirements, must be reconstructed
within $|\eta|<2.47$ (excluding the barrel-endcap transition region $1.37 < |\eta| < 1.52$), and have at least 20~\gev of \pt~\cite{EGAM-2018-01}.

Muons are reconstructed by combining a track in the inner tracking detector with either a track or hits in the muon spectrometer.  The ‘loose’ quality identification and isolation requirements are imposed, together with kinematic requirements of $|\eta|<2.4$ and $\pt > 20~\gev$~\cite{MUON-2018-03}.

Small-$R$ jets, reconstructed from PFOs using the anti-$k_t$ algorithm with a radius parameter of $R=0.4$ are also used, either to define a recoiling reference system or to veto additional radiation.
The small-$R$ jets are calibrated using a set of simulation and \emph{in situ} techniques similar to the ones employed for the large-$R$ jets of this paper~\cite{JETM-2018-05}.
Only small-$R$ jets with  $|\eta|<2.5$ and $\pt > 10~\gev$ are considered.
Small-$R$ jets with $\pt<60~\gev$ are required to satisfy basic cleaning requirements and the jet vertex tagger (JVT) selection~\cite{PERF-2014-03} to suppress pile-up contamination. The \enquote{tight} JVT working point is used for the nominal selection and the \enquote{medium} variation is used to evaluate the associated systematic uncertainty.
All small-$R$ jets within $\Delta R = 0.35$ of a lepton or $\Delta R = 0.2$ of a photon are removed.
Only small-$R$ jets with $\Delta R>1.4$ to the large-$R$ jets used to measure the large-$R$ jet energy detector response are considered.

\subsection{Pseudorapidity intercalibration}
\label{sec:eta_intercalibration}
The relative $\eta$-intercalibration extends the large-$R$ UFO jet calibration from the central detector to the forward region, $0.8 < |\eta_\text{det}| < 2.5$, which is typically known with less precision due to the more complicated detector structure.
This calibration is derived following the so called \emph{matrix~method}~\cite{JETM-2022-01}, by studying in data and simulations the difference between the \pt balance between pairs of large-$R$ UFO jets produced in different $\eta_\text{det}$ regions.
This technique outperforms previously used ones~\cite{JETM-2018-05} as it increases the statistical precision of the measurements by removing the requirement for one of the two large-$R$ UFO jets to be in the central reference region:
instead, all regions are calibrated relative to one another by solving a set of linear equations.

The asymmetry between the large-$R$ UFO jet produced with lower $\eta_\text{det}$ values (\enquote{left} jet) and the one produced with higher $\eta_\text{det}$ (\enquote{right} jet), defined as $\mathcal{A}=(\pt^\text{left}-\pt^\text{right})/ \pt^\text{avg}$, is studied in bins of the average transverse momentum $\pt^\text{avg}=(\pt^\text{left}+\pt^\text{right})/2$ and $|\eta_\text{det}|$ to obtain the large-$R$ UFO jet energy response as explained in Ref.~\cite{JETM-2018-05}.

Multijet events in data are selected using several dedicated single-jet triggers.
Their efficiency is evaluated for large-$R$ UFO jets and each trigger is used in its region of maximum efficiency for those jets.
These triggers allow for a study of the detector response of large-$R$ UFO jets with \pt from 180~\gev to 3~\tev.
To ensure a back-to-back dijet topology, events with energetic additional radiation are vetoed by an upper limit on the transverse momentum of the \pt-leading small-$R$ jet not geometrically matched to the selected large-$R$ UFO jets ($\pt^{j}$), and the two leading large-$R$ UFO jets are required to satisfy a minimum azimuthal angle separation.
Both of these requirements are varied to evaluate their impact on the response measurements by taking $\pt^{j}/\pt^\text{avg}<0.25\pm0.1$ and $\Delta\phi>2.5\pm0.3$.
The observed differences between the results are taken as uncertainties.
The selection requirement on the small-$R$ jet is different from what was used in Ref.~\cite{JETM-2018-02}, where a large-$R$ jet was used instead. The updated requirements allow for a more precise vetoing of soft radiation spoiling the dijet topology.

\begin{figure}[tp]
\centering
\subfloat[]{\includegraphics[width=0.49\linewidth]{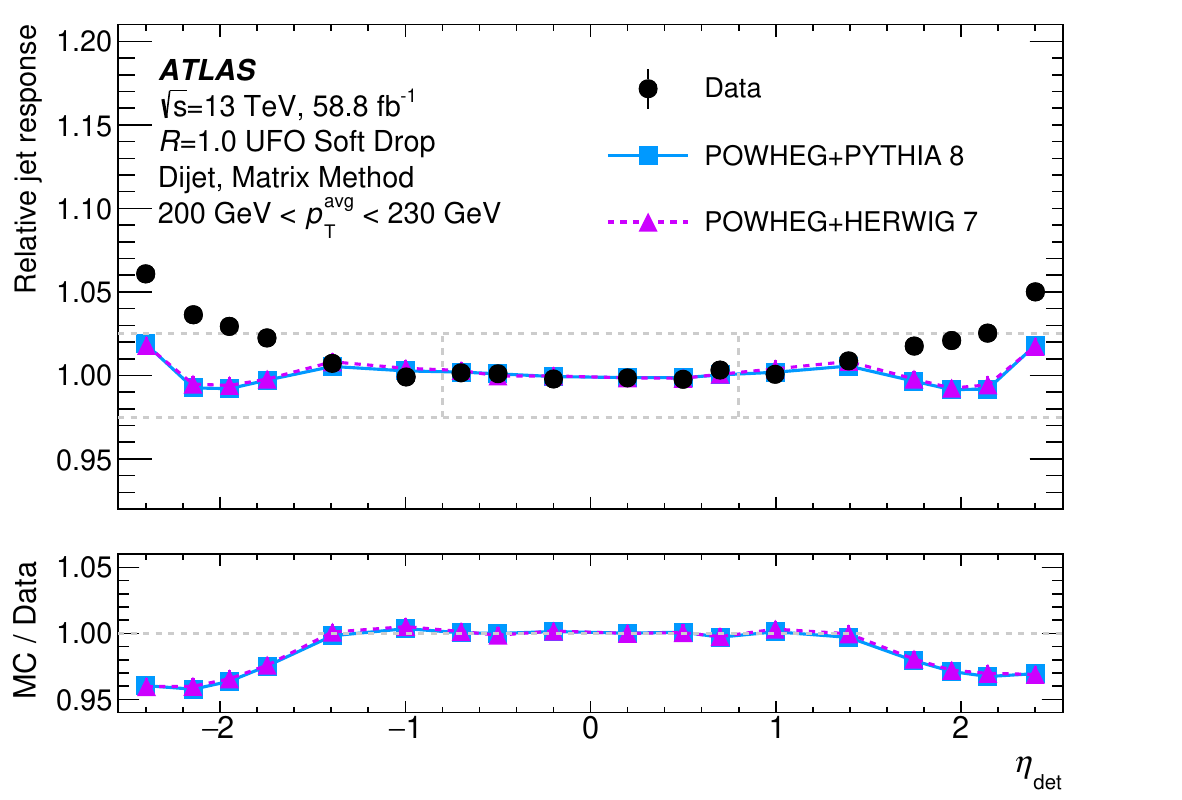}}
\subfloat[]{\includegraphics[width=0.49\linewidth]{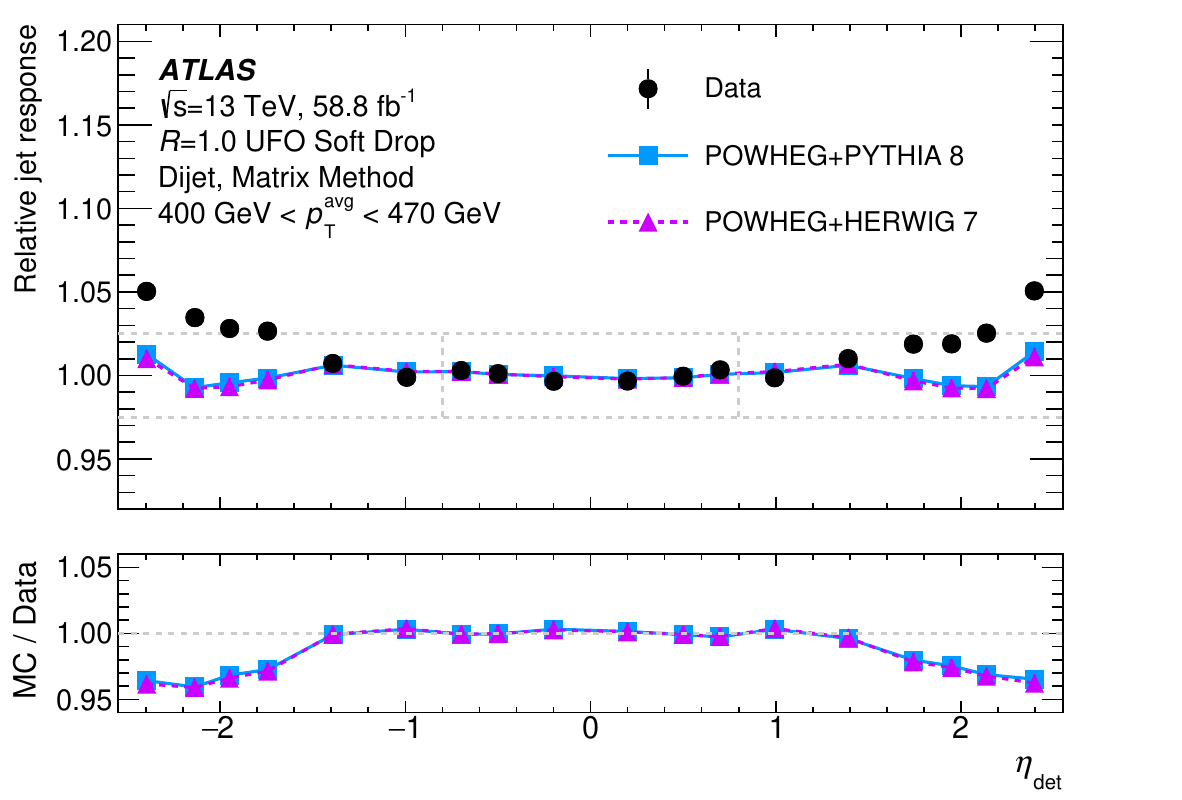}}
\caption{The relative large-$R$ UFO jet response obtained with the matrix method in dijet events against the detector pseudorapidity, for data collected in 2018 with Monte Carlo predictions from \POWPY~8 and \POWHER~7, for (a) $200~\GeV<p_\text{T}^\text{avg}<230~\GeV$ and (b) $400~\GeV<p_\text{T}^\text{avg}<470~\GeV$. Jet response is obtained relative to the central detector region ($|\eta_\text{det}|<0.8$), indicated by vertical dotted lines. The lower panel shows the ratio of the results obtained with Monte Carlo simulation and data. Statistical uncertainties are represented by vertical lines, but are too small to be visible.%
}
\label{fig:etainter_factors}
\end{figure}

The relative large-$R$ UFO jet \pT response for 2018 data is shown in Figure~\ref{fig:etainter_factors} for data, \POWPY~8 and \POWHER~7.  The relative response in data increases in the forward region due to features of the experimental response that are not well-reproduced in the simulation and hence not accounted for in the simulation-based JES calibration factor.

\begin{figure}[tp]
\centering
\subfloat[]{ \includegraphics[width=0.49\linewidth]{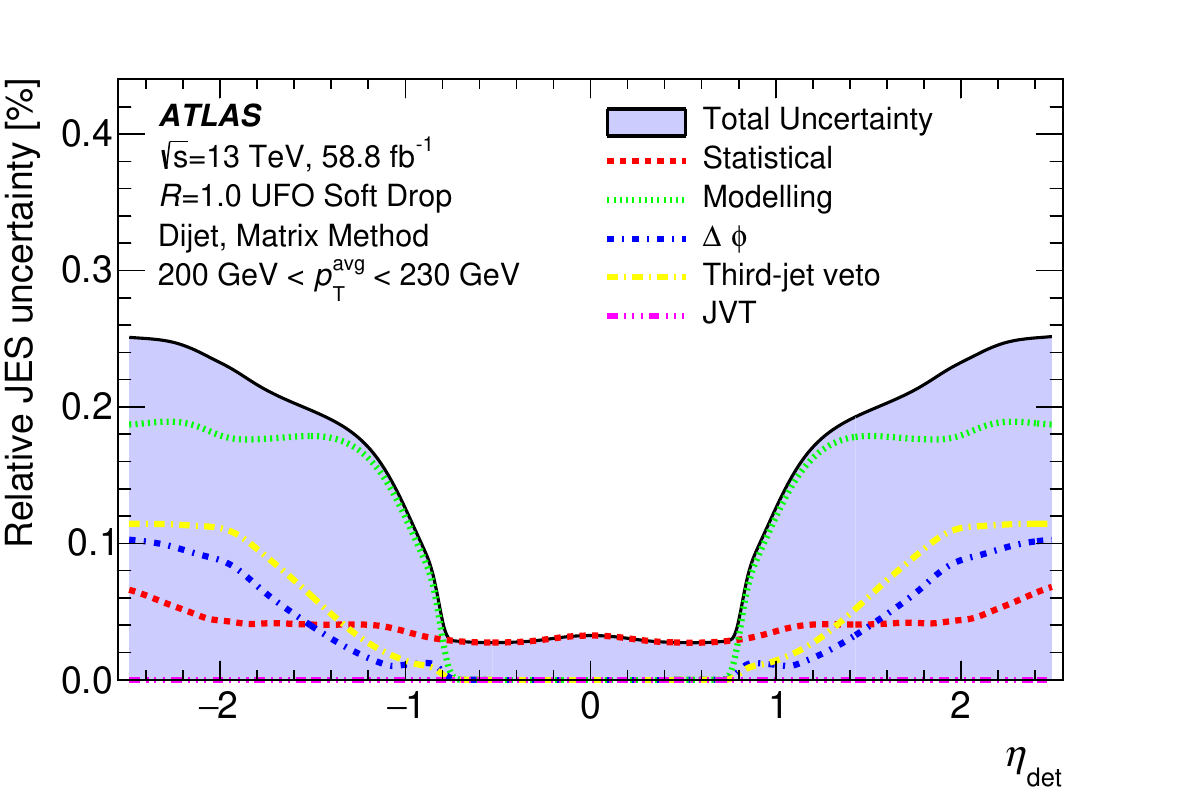}}
\subfloat[]{ \includegraphics[width=0.49\linewidth]{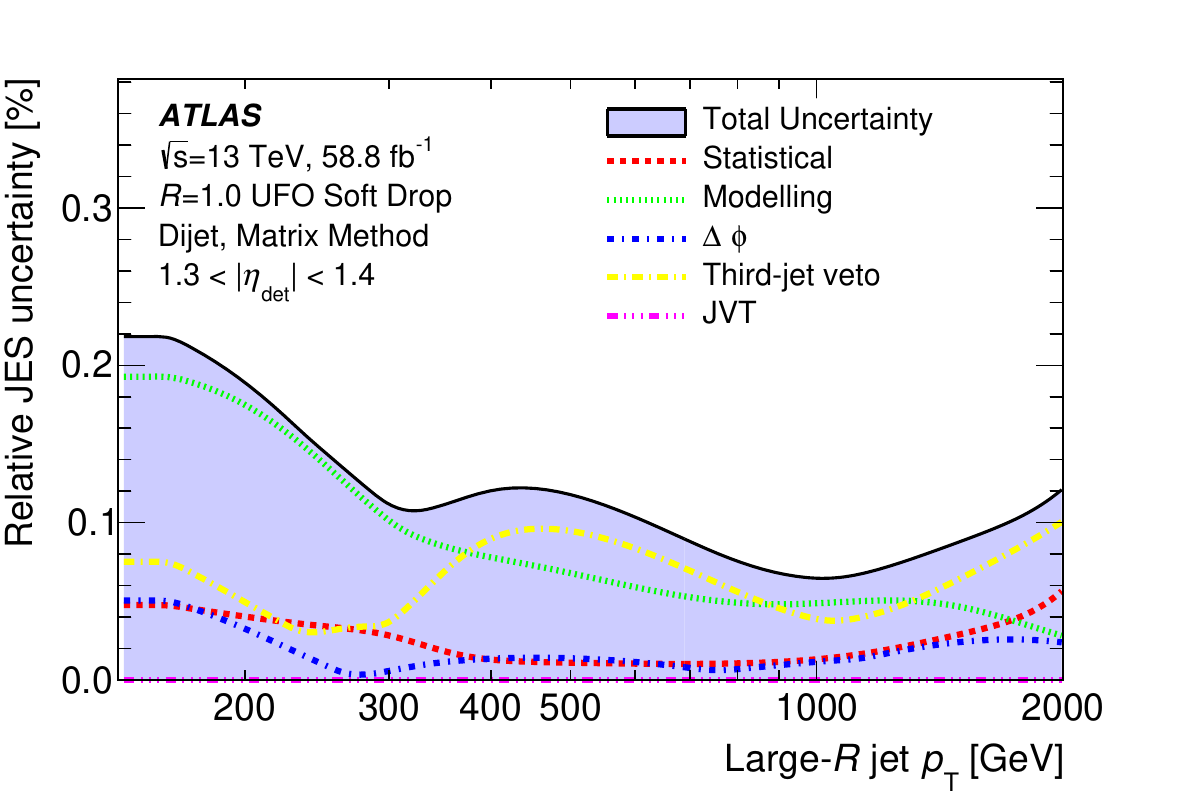}}
\caption{The breakdown of uncertainties affecting the relative large-$R$ UFO jet energy response with the matrix method as a function of (a) the detector pseudorapidity and (b) the large-$R$ UFO jet \pt. The uncertainties are shown for the 2018 data sample, with other years showing a compatible behaviour. }
\label{fig:etainter_uncs}
\end{figure}

Figure~\ref{fig:etainter_uncs} shows the uncertainties affecting the response measurement. The multijet modelling uncertainty obtained by comparing the \POWPY~8 and \POWHER~7 simulations is the largest source of uncertainties at 0.2\% for $\pT<300~\GeV$, while falling to about 0.05\% for higher \pT values.
Systematic uncertainties related to the event selection criteria are typically sub-dominant at low-\pT, however the small-$R$ jet veto becomes the largest uncertainty at  around 0.1\% for \pT values above 400~\GeV. The relative responses obtained for different data years and their associated uncertainties are found to be compatible with each other.

\FloatBarrier

\subsection{$Z$+jet balance method}
\label{sec:zjets}

The large-$R$ UFO jet energy response is studied in collision data with the direct balance method~\cite{JETM-2018-02} in $Z$+jets events, where the $Z$ boson decays to a pair of electrons or muons.  The average transverse momentum balance between the large-$R$ UFO jet and the $Z$ boson is defined as
$R_{Z+\text{jets}}=\left\langle\pt^\text{jet}/\pt^\text{ref}\right\rangle$,  where $\pt^\text{jet}$ is the large-$R$ jet \pt and $\pt^\text{ref}=\pt^{Z}|\cos{\Delta\phi}|$ is the component of the reconstructed $Z$-boson momentum projected onto the direction of the large-$R$ UFO jet in the transverse plane, with $\Delta\phi$ being the azimuthal angle between the large-$R$ UFO jet and reference $Z$ boson. %

Selected events are required to satisfy a di-electron or di-muon trigger, which are fully efficient for leptons with $\pt>20~\gev$~\cite{TRIG-2018-01,TRIG-2018-05,TRIG-2019-03,TRIG-2016-01}.
To enhance the probability of the lepton pair coming from the decay of a $Z$ boson, the leptons of the $\ell\ell$ system are required to have opposite electric charges and an invariant mass $66~\gev < m(\ell^+\ell^-) < 116~\gev$.

Only large-$R$ UFO jets in the central region of the detector ($|\eta_\text{det}|<0.8$) and with $\pt>150~\gev$ are considered.
A requirement of $\Delta\phi>2.8$ improves the transverse momentum balance between the large-$R$ UFO jet and the $Z$ boson. Additional event activity is suppressed by rejecting  events in which any small-$R$ jet  is reconstructed with a $\pT > \text{max}(0.1 \times \pt^\text{ref}, 15~\gev)$. These selections are varied as  $\Delta\phi>2.8\pm0.1$ and  $\pT > \text{max}( [0.1\pm 0.05] \times \pt^\text{ref}, 15 \pm 5 ~\gev )$ to estimate the impact of the event selection choices on the detector response measurement.

\begin{figure}[!htp]
\centering
\subfloat[]{ \includegraphics[width=0.49\linewidth]{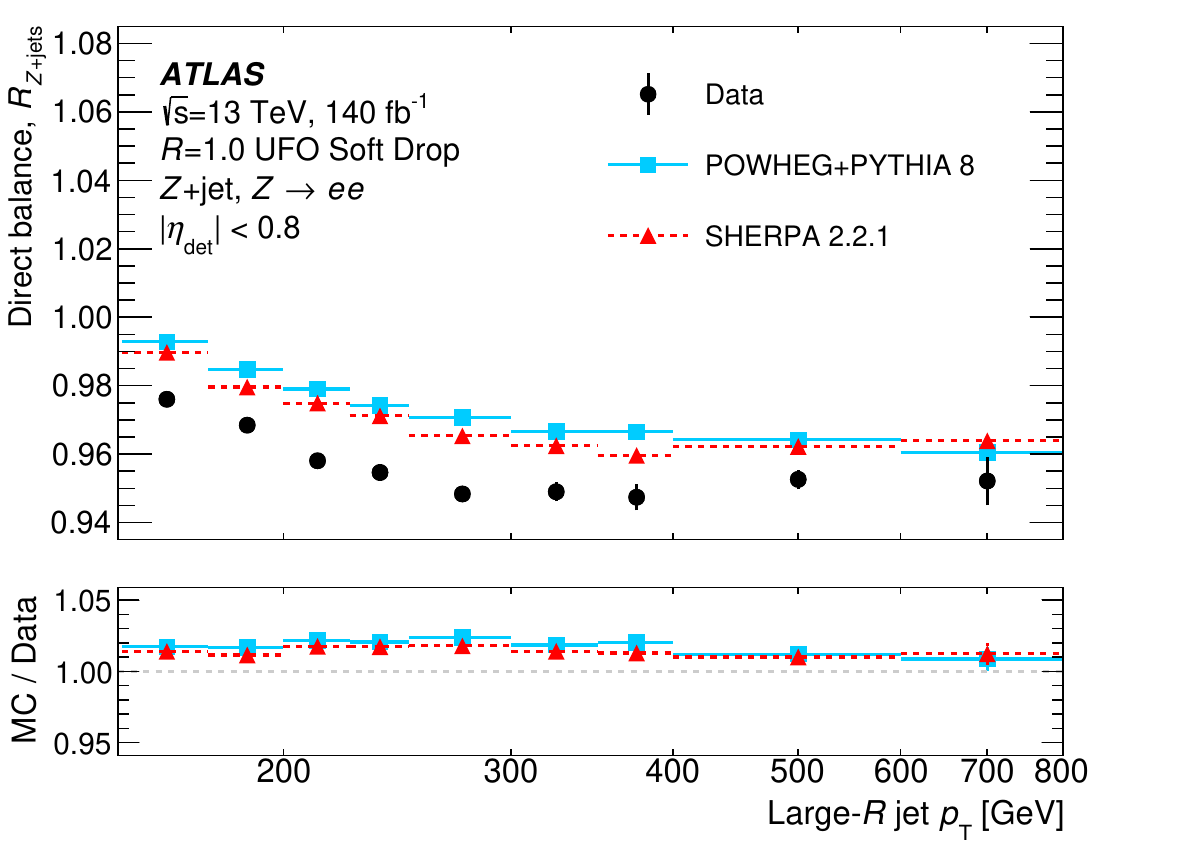}}
\subfloat[]{ \includegraphics[width=0.49\linewidth]{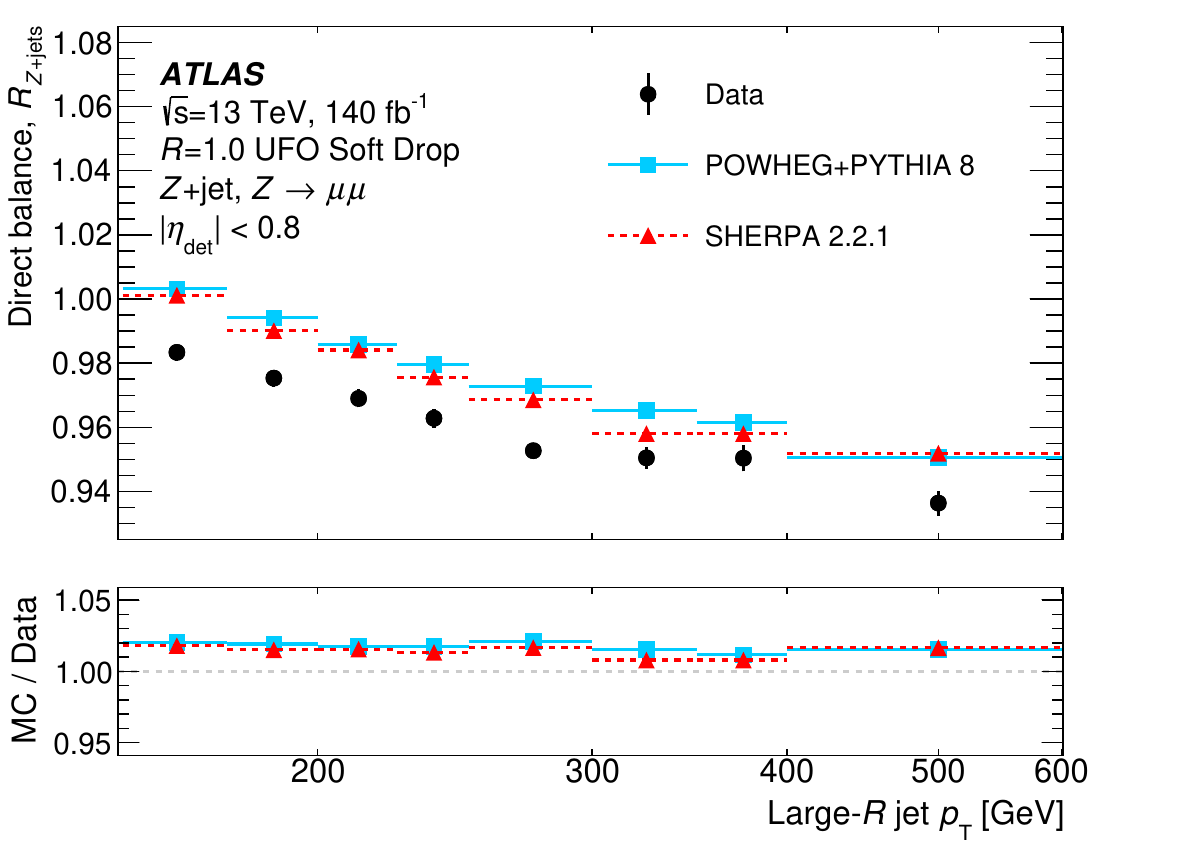}}
\caption{The transverse momentum balance in (a) $Z(\rightarrow ee)$+jets  and (b) $Z(\rightarrow\mu\mu)$+jets events as a function of the large-$R$ UFO jet \pt. The results obtained on the full Run~2 data are compared with expectations from \POWPY and \SHERPA. The lower panel shows the ratio of the results obtained with Monte Carlo simulation and data. Statistical uncertainties are represented by vertical lines.}
\label{fig:zjet_rdb}
\end{figure}

Figure~\ref{fig:zjet_rdb} shows the $R_{Z+\text{jets}}$ response for data and for two MC samples using different generators, \POWPY[8] and \SHERPA[2.2.1], in the $Z\rightarrow e^+e^-$ and $Z\rightarrow \mu^+\mu^-$ channels.
The balance values are found to range between 1 and 0.95, closer to unity than what was obtained for large-$R$ calorimeter jets~\cite{JETM-2018-02}%
.
The average balance is found to be 1.5 -- 2.0\% lower in data than in the MC simulation, in both $Z$-boson decay channels and across the \pt{} range accessible with this method.

\begin{figure}[!ht]
\centering
\subfloat[]{ \includegraphics[width=0.49\linewidth]{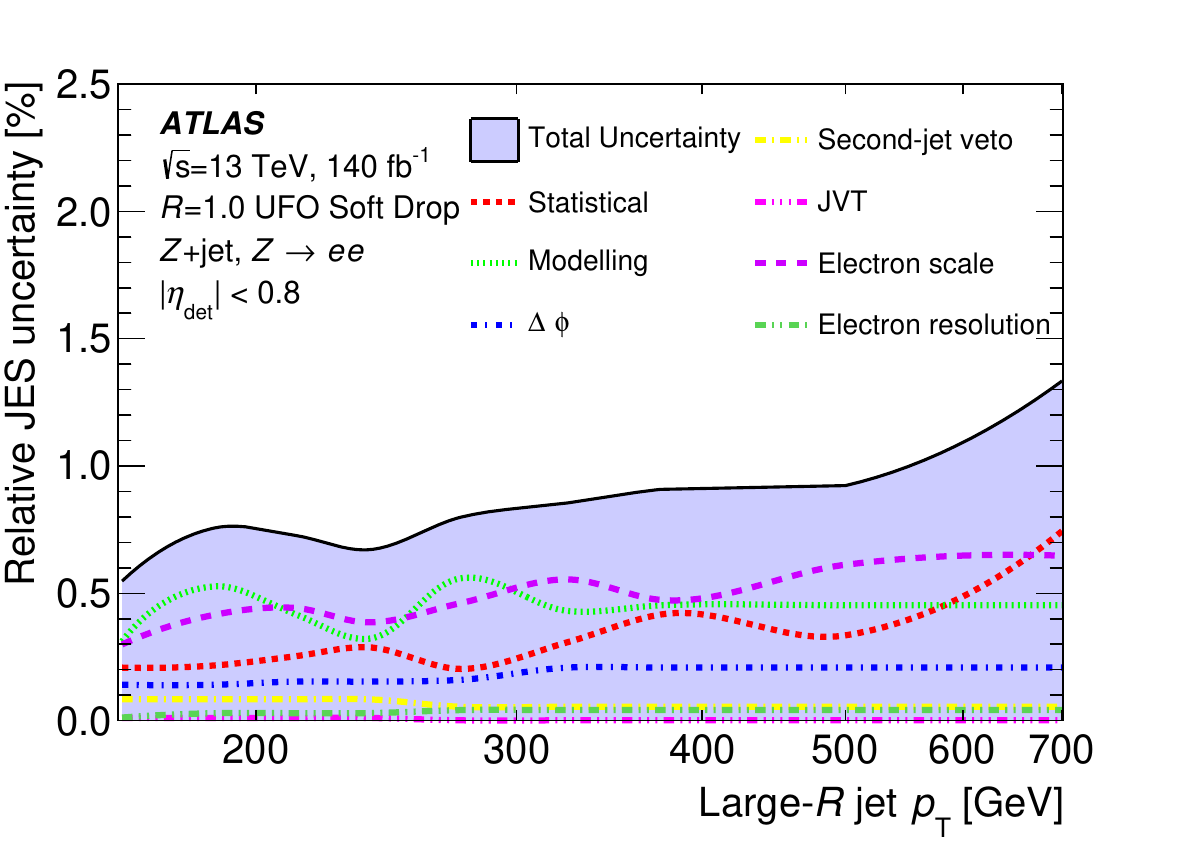}}
\subfloat[]{ \includegraphics[width=0.49\linewidth]{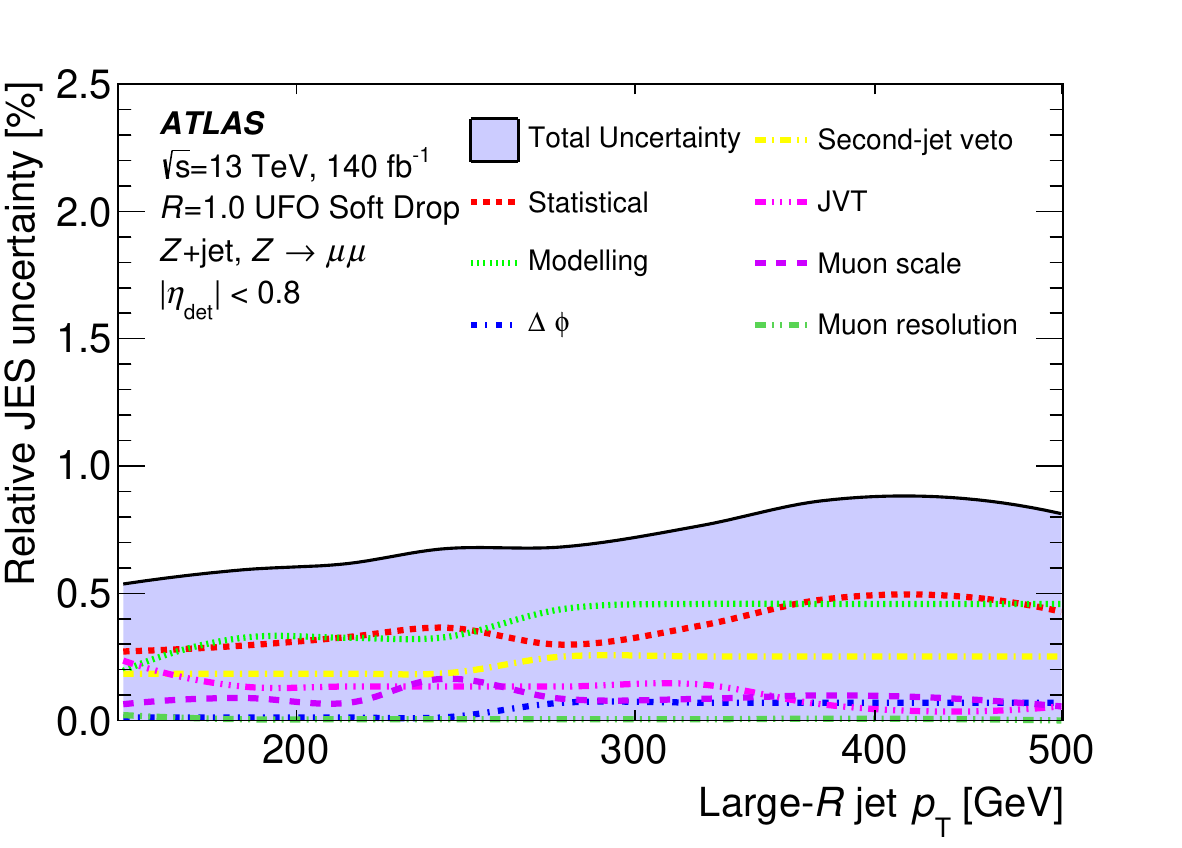}}
\caption{The breakdown of uncertainties affecting the relative large-$R$ UFO jet energy response with the (a) $Z(\rightarrow ee)$+jets  and (b) $Z(\rightarrow\mu\mu)$+jets balance method, as a function of the leading large-$R$ UFO jet \pt.}

\label{fig:zjet_uncs}
\end{figure}

The main uncertainties affecting the detector response measurement are found to be the MC modelling (both channels, obtained by comparing \POWPY[8] and \SHERPA[2.2.1] results), the electron energy scale uncertainty (for the electron channel only) and the statistical uncertainty of the data. Other sources of uncertainty are found to be less relevant, as shown in Figure~\ref{fig:zjet_uncs}. The total uncertainty in the \emph{in situ} correction factor increases from 0.5\% at low \pt{} to slightly over 1\% at $\pt = $ 750~\gev, in both channels.

The $Z$+jet direct balance method provides a precise measurement of the large-$R$ UFO transverse momentum response for jets with transverse momenta between 150 and 750~\gev{} and is an important component of the ATLAS \emph{in situ} jet calibration in this range. Above 750~\gev, the number of the selected events do not allow to obtain a precise determination and other event topologies are studied.
Compared with the previous uncertainty in the $Z$+jet \emph{in situ} large-$R$ calorimeter jet calibration for 2015--2016~\cite{JETM-2018-02}, modelling uncertainties are now slightly reduced and the measurement is extended to a higher \pt range.

\FloatBarrier

\subsection{$\gamma$+jet balance method}
\label{sec:gammajets}
The balance method in $\gamma$+jets events follows the same methodology as in Section~\ref{sec:zjets}, where a photon is used instead of a $Z$ boson.
Events are selected using the lowest unprescaled single-photon trigger with a threshold of 140~\GeV. Offline selection requires a photon that satisfies the \enquote{tight} identification and isolation criteria~\cite{EGAM-2021-01} and has $E_\text{T} \geq 145$~\GeV to ensure a trigger efficiency above 99\%. Only photons with $|\eta|<1.37$ are considered to ensure they have the most precise energy scale determination~\cite{EGAM-2021-02}. Significant additional radiation affects the assumption of a balanced event topology, so events are vetoed if a reconstructed, calibrated small-$R$ jet has $p_\text{T} > \text{max}(0.1 \times \pt^\text{ref}, 15~\text{GeV})$.
To ensure back-to-back production of photon and a large-$R$ UFO jet, photons are required to be separated from reconstructed large-$R$ UFO jets by at least $\Delta \phi (\gamma,\text{large-$R$ jet}) > 2.8$. To assess the impact on the measurement of these requirements, the selections are varied similarly to what is done in Section~\ref{sec:zjets}. Additionally, the impact of the requirement vetoing an overlap between small-$R$ and large-$R$ UFO jets is varied as $\Delta R=1.4\pm0.2$.

A photon purity correction is applied to the average results of the balance in data to account for contamination from misidentified jets or electrons, which can distort the nominal $p_\text{T}$ balance. This correction is implemented as a statistical weight applied to the event as a function of the photon \pt. The contamination from fake photons in the photon sample is estimated by using the double-sideband (ABCD) method, which relies on a plane defined by information on photon isolation and photon identification. This method assumes that isolation and identification are uncorrelated, which enables the calculation of the probability of real photons in the simulation to correct for background effects. The purity correction results in a shift in the $R_{\gamma\text{+jets}}$ value in data at the per-mil level.

\begin{figure}[!htp]
\centering
\includegraphics[width=0.6\linewidth]{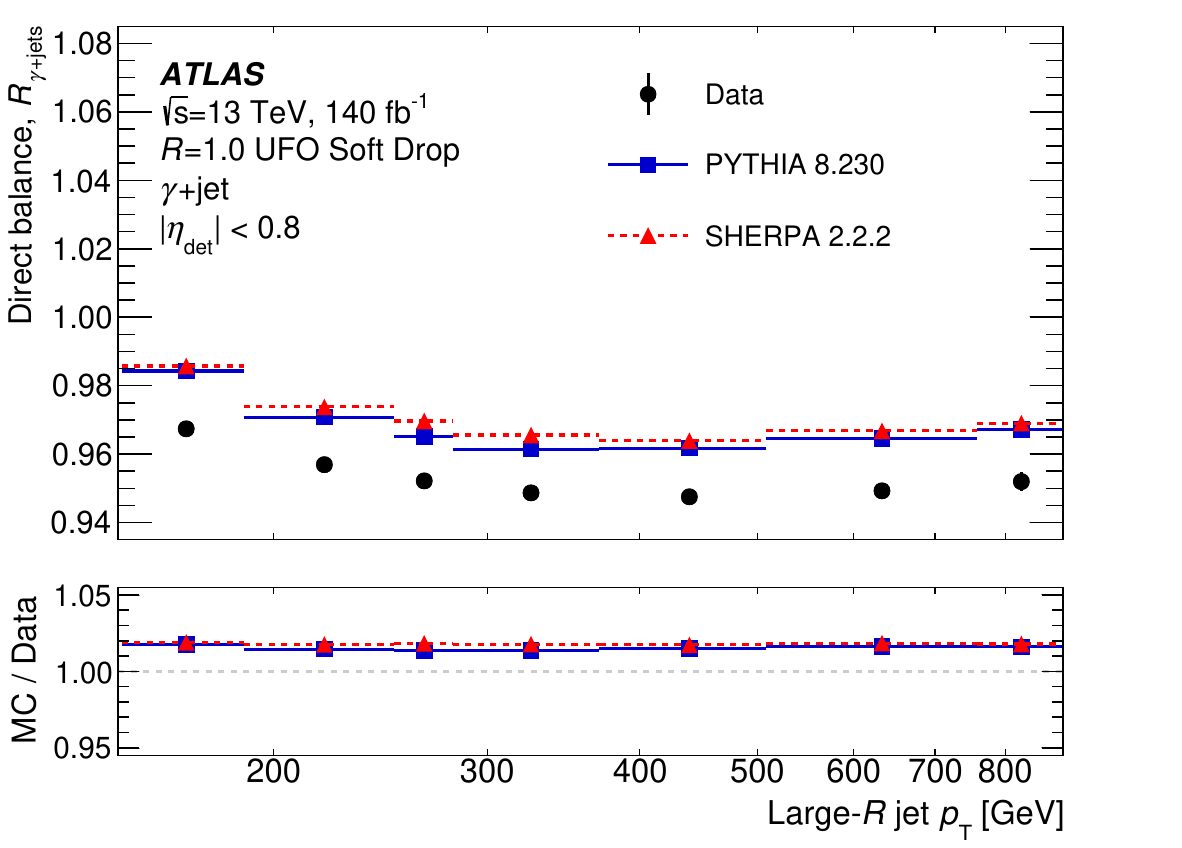}
\caption{The transverse momentum balance in $\gamma+$jets events as a function of the large-$R$ jet \pt. The results obtained on the full Run~2 data are compared with expectations from \PYTHIA and \SHERPA.  The lower panel shows the ratio of the results obtained with Monte Carlo simulation and data. Statistical uncertainties are represented by vertical lines, but are too small to be visible.}%
\label{fig:gamma_jet_balance}
\end{figure}

The direct balance result is shown in Figure~\ref{fig:gamma_jet_balance} as a function of the large-$R$ UFO jet $p_\text{T}$ in the central region of the detector ($|\eta_\text{det}| < 0.8$). The bottom panel compares the predicted response from the \PYTHIA and \SHERPA simulations to the response observed in data, with the MC-to-data ratio exceeding unity across the entire momentum range. This ratio serves as the correction factor that is applied to the data for \emph{in situ} corrections.

\begin{figure}[!htp]
\centering
\includegraphics[width=0.6\linewidth]{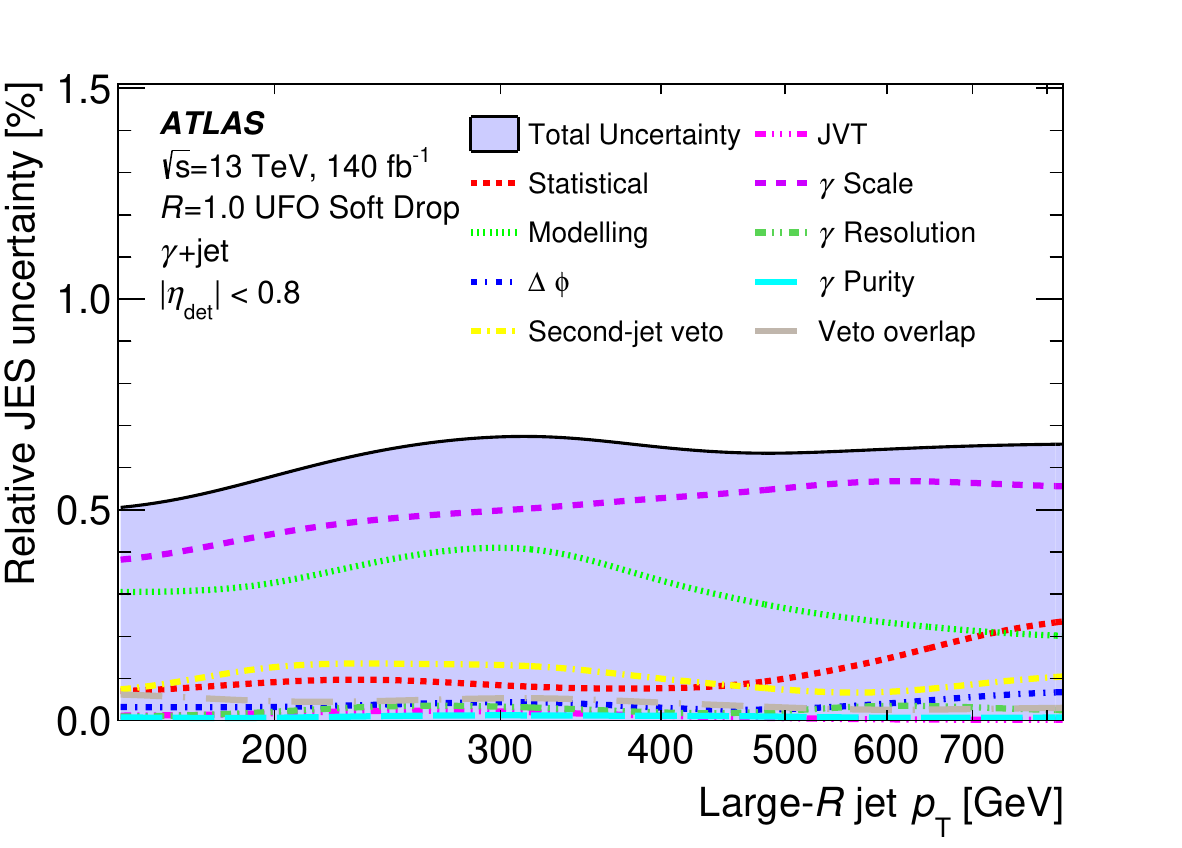}
\caption{The breakdown of uncertainties affecting the relative large-$R$ UFO jet energy response with the $\gamma+$jets balance method, as a function of the leading large-$R$ UFO jet transverse momentum.}
\label{fig:gamma_jet_uncertainty}
\end{figure}

As shown in Figure~\ref{fig:gamma_jet_uncertainty}, the overall statistical and systematic uncertainties for the \emph{in situ} measurement remain below 0.8\% across the entire large-$R$ UFO jet $p_\text{T}$ range, with an average uncertainty of approximately 0.7\%. The dominant source of uncertainty throughout is the photon energy scale. Compared with the previous uncertainty in the $\gamma$+jet \emph{in situ} large-$R$ calorimeter jet calibration for 2015--2016~\cite{JETM-2018-02}, this represents a decrease in uncertainty of roughly 40\%. This improvement is largely due to advancements in the photon energy scale calibration. The photon energy scale uncertainty in the central region of the detector, for the photons used in this analysis, has halved compared with its estimate in 2015--2016~\cite{PERF-2017-03}, due to a more refined assessment of systematic uncertainties in the electromagnetic calorimeter response~\cite{EGAM-2021-02}, in particular in the gain calibration of layer~2 and in the inter-calibration of layers~1 and~2.
The second most significant uncertainty is from MC modelling, which is calculated as the difference between response between \PYTHIA and \SHERPA, both used for measuring the direct balance. This uncertainty has improved compared with previous results, thanks to upgrades in both MC generator versions. The transition from \SHERPA 2.1.1 to 2.2.2 and from \PYTHIA 8.1 to 8.2 has brought the two event generators into closer agreement in their treatment of photon+jet processes. These newer versions include more refined treatment of QCD corrections and photon radiation. Both tools have also improved their parton shower algorithms and phase space sampling, resulting in more accurate simulations of $\gamma$+jet events~\cite{Sjostrand:2014zea, Bothmann:2019yzt}.

The photon-jet balance method provides good precision for large-$R$ UFO jet transverse momenta up to 800~\gev. The recommendations based on the full Run~2 photon energy scale uncertainties and newer versions of the MC generators lead to an important reduction in the main systematic uncertainties compared with the large-$R$ calorimeter jet energy scale determination with this method based on 2015 and 2016 data.

\FloatBarrier

\subsection{Multijet balance method}
\label{sec:multijetbalance}
The energy scale of higher-\pt large-$R$ UFO jets is measured using multijet events, through the multijet balance (MJB) method.
Events are selected where an energetic large-$R$ UFO jet is balanced against a system that consists of multiple lower-\pT fully calibrated small-$R$ jets. The transverse momentum $\pt^\text{recoil}$ of the recoil system is obtained as the four-vector sum of the calibrated small-$R$ jets.
The transverse momentum balance $R_\text{MJB} =\left\langle \pt^\text{jet}/\pt^\text{recoil} \right\rangle$ is measured in data and determined in simulation in several bins of $\pt^\text{recoil}$.

Events are selected using single small-$R$ jet triggers. Bins of $\pt^\text{recoil}$ are defined to correspond to a given fully efficient single small-$R$ jet trigger.
The triggers used for $200~\gev <\pt^\text{recoil}< 550~\gev$ are prescaled, whereas an un-prescaled jet trigger is used for $\pt^\text{recoil}> 550~\gev$.

The large-$R$ UFO jet is required to have $|\eta_\text{det}|< 0.8$ and at least three small-$R$ jets with $|\eta|< 2.8$ and $\pt > 25~\gev$ are required to build the recoil system.
To enhance the multijet topology and ensure that the recoil system consists of several well-calibrated small-$R$ jets with lower \pt{} than the large-$R$ UFO jet, the leading small-$R$ jet in the recoil system is allowed to have no more than 80\% of the total transverse momentum of the recoil system.
The angle $\phi$ in the azimuthal plane between the leading large-$R$ UFO jet and the vector defining the recoil system is required to satisfy $|\phi - \pi|<0.3$, while the angular distance $\Delta R$ between the leading large-$R$ UFO jet and the nearest small-$R$ jet from the recoil system is required to be greater than 1.5.
Such kinematic requirements are varied to estimate systematic uncertainties associated with the event selection by varying the transverse momentum
threshold for recoil small-$R$ jets by $\pm5$~\GeV, the percentage of total recoil transverse momentum carried by the leading recoiling small-$R$ jet by $\pm10\%$, the requirement on the angle $\phi$ by $\pm 0.1$, and the selection on $\Delta R$ by $\pm0.5$. This last variation was modified relative to the $\pm0.4$ variation used in Ref.~\cite{JETM-2018-02}.

\begin{figure}[!htp]
\centering
\subfloat{ \includegraphics[width=0.65\linewidth]{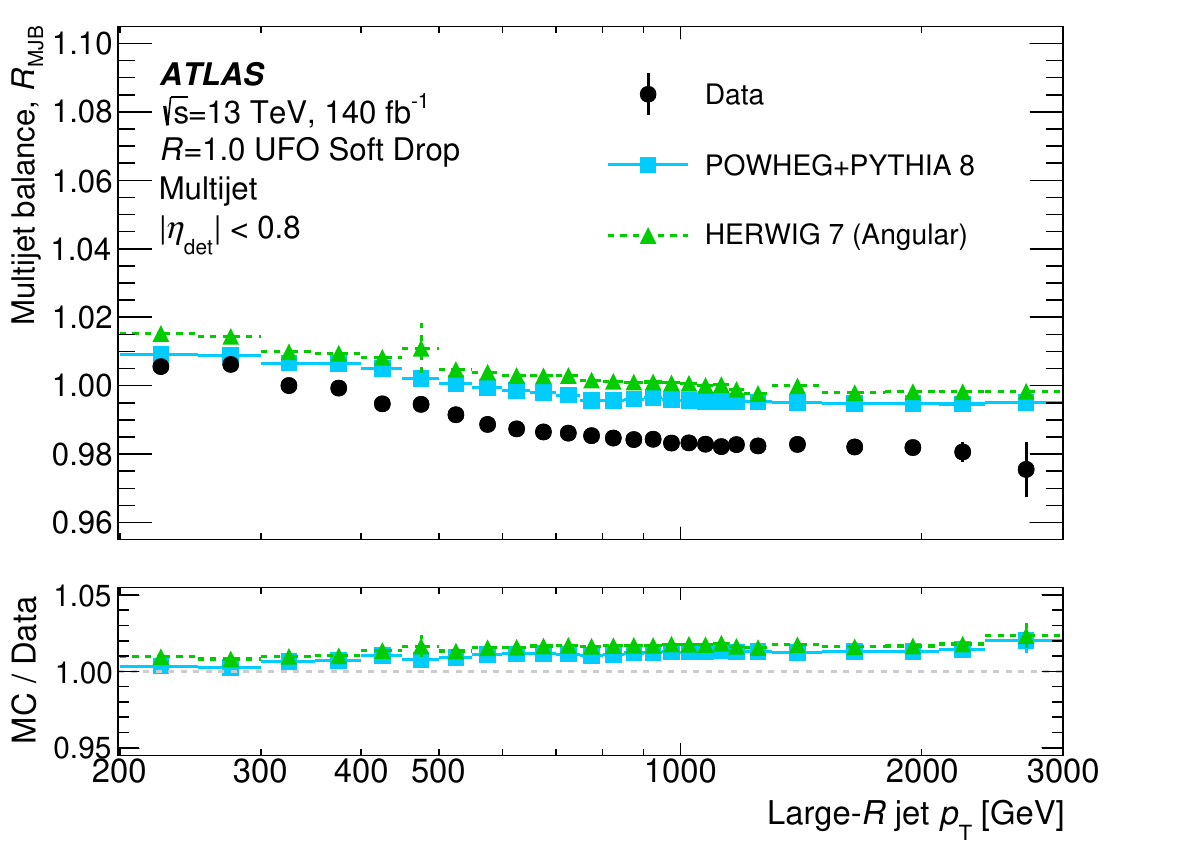}}
\caption{ The transverse momentum balance of the leading large-$R$ UFO jet and the recoiling system of multiple small-$R$ jets, as a function of the large-$R$ UFO jet \pt.
The data are compared with the \POWPY nominal prediction and to a sample generated with \HERWIG using angular ordering in the parton shower, which is found to yield the largest difference in the balance to the nominal prediction among various other multijet models tested.
The lower panel shows the ratio of the results obtained with Monte Carlo simulation and data.
Statistical uncertainties are represented by vertical lines.}
\label{fig:mjb_calib}
\end{figure}

The $R_\text{MJB}$ values for data, \POWPY[8] and \HERWIG angular-ordered are shown in Figure~\ref{fig:mjb_calib}.

\begin{figure}[!htp]
\centering
\includegraphics[width=0.65\textwidth]{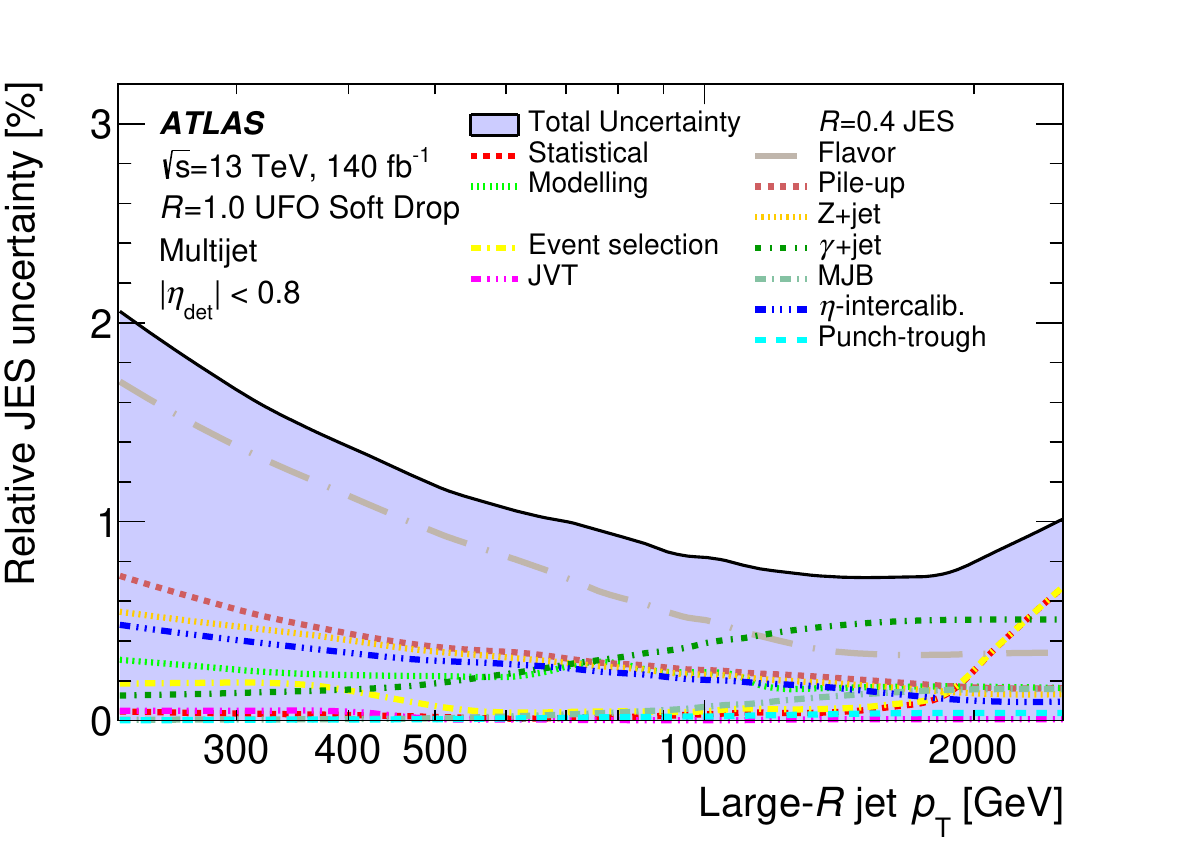}
\caption{The breakdown of uncertainties affecting the relative large-$R$ UFO jet energy response with the multijet balance method, as a function of the leading large-$R$ UFO jet \pt. }
\label{fig:mjb_unc}
\end{figure}

The uncertainties affecting the detector response of the multijet balance method are summarised in Figure~\ref{fig:mjb_unc}.
The uncertainty in the lower \pt range is dominated by the flavour dependence of the small-$R$ JES response, while at higher \pT the statistical uncertainty and the component associated with the small-$R$ JES measurement in $\gamma$+jet events become important.
As opposed to the $R_{Z+\text{jets}}$ and $R_{\gamma+\text{jets}}$ measurements, here the dependence on the MC modelling is not very relevant. The MC modelling systematic uncertainty is defined as the maximum distance to \POWPY[8] of the \HERWIG generator with angular-ordered parton showering, which was found to give the largest difference in response to \POWPY[8] among the multijet MC generators introduced in Section~\ref{sec:MC}.

The multijet balance method provides a precision of approximately 1\% over a broad \pt{} range from 500~\gev{} to 3~\tev{} and is crucial to expand the coverage of the \emph{in situ} calibration to the high-\pt{} regime.

\FloatBarrier


\section{\emph{In situ} jet energy resolution}
\label{sec:insitujer}
%

%
%
%
%
%

The resolution of large-$R$ UFO jets \pt is measured \emph{in situ} in multijet events, by studying the momentum balance between two large-$R$ UFO jets.
The same event selection criteria and trigger strategy are used as in the $\eta$-intercalibration described in Section~\ref{sec:eta_intercalibration}.
Additionally, large-$R$ UFO jets studied in this section are also calibrated using the corrections derived from the MC calibrations and \emph{in situ} energy calibrations described in Sections~\ref{sec:mccalib} and~\ref{sec:jes_insitu}.

The width of the asymmetry distribution $\mathcal{A}=(\pt^\text{jet1}-\pt^\text{jet2})/\pt^\text{avg}$ is studied, as it depends on the resolution of the jet \pt measurement and on the intrinsic particle-level width, which arises due to balance fluctuations and out-of-cone effects.
Since the latter effect is uncorrelated with the detector response, the component of the asymmetry width due to the detector resolution, $\sigma_{\mathcal{A}\text{, det}}$, can be determined by subtracting in quadrature the
asymmetry width of particle level jets, $\sigma_{\mathcal{A}\text{, truth}}$,  from that of reconstructed jets, $\sigma_{\mathcal{A}\text{, reco}}$.
The asymmetry distribution is studied in bins of $\pt^\text{avg}$ and
in different $\eta_\text{det}$ bins: a reference region with $0.2<|\eta_\text{det}|< 0.7$ (\enquote{ref} region), where the seam at |$\eta_\text{det}| = 0$ is excluded to ensure the reference jet energy is as cleanly measured as possible,  and \enquote{probe} regions covering the rest of the phase space up to $|\eta_\text{det}| < 2.0$. Six different probe regions are considered. %

In the context of a well-balanced dijet topology, if both large-$R$ UFO jets are within the central reference region they have the same \pt resolution.
Since both large-$R$ UFO jets contribute the same amount to the asymmetry distribution, the relative jet-\pT resolution of the reference region is obtained as
\begin{equation}
\label{eqn:JER_probeinref}
\left(\frac{\sigma_{\pt}}{\pt}\right)_\text{ref}=\frac{\sigma_{\mathcal{A}\text{, det}}^\text{ref}}{\sqrt{2}}\quad .
\end{equation}
If one large-$R$ UFO jet lies in a probe region, its \pt resolution can be extracted from the width of the asymmetry distribution of the large-$R$ UFO jets in the probe and central region as
\begin{equation}
\label{eqn:JER_probeoutsideref}
\left(\frac{\sigma_{\pt}}{\pt}\right)_\text{probe}=\sqrt{ \left(\sigma_{\mathcal{A}\text{, det}}^\text{probe}\right)^2 - \left(\frac{\sigma_{\mathcal{A}\text{, det}}^\text{ref}}{{\sqrt{2}}}\right)^2}\quad .
\end{equation}
These formulas assume no correlation between the momentum resolutions of the two large-$R$ UFO jets.
It is observed that there is a non-zero correlation, $\rho$, which makes the former formulas approximate.
In presence of a correlation term, Equations~\ref{eqn:JER_probeinref} and~\ref{eqn:JER_probeoutsideref} are modified into:
\begin{equation*}
\left(\frac{\sigma_{\pt}}{\pt}\right)_\text{ref}=\frac{\sigma_{\mathcal{A}\text{, det}}^\text{ref}}{\sqrt{2}}\times\frac{1}{\sqrt{1-\rho}}.
\end{equation*}
and
\begin{equation*}
\left(\frac{\sigma_{\pt}}{\pt}\right)_\text{probe}=\frac{\sigma_{\mathcal{A}\text{, det}}^\text{ref}}{\sqrt{2}}\times\frac{\rho}{\sqrt{1-\rho}} + \sqrt{ \left(\sigma_{\mathcal{A}\text{, det}}^\text{probe}\right)^2 + \frac{(\rho^2 -1)}{2\sqrt{1-\rho}}\times \left(\sigma_{\mathcal{A}\text{, det}}^\text{ref}\right)^2 },
\end{equation*}
respectively.
The values of $\rho$ are obtained in simulation in each $|\eta_\text{det}|$ and $\pt^\text{avg}$ bin, using the distributions of large-$R$ UFO jet $\pt$ minus the matched particle-level jet $\pt$,  $\Delta_{j_{i}} \equiv p_{\text{T}_{j_{i}}}^{\text{reco}} - p_{\text{T}_{j_{i}}}^{\text{truth}}$, where $i=1,2$ is the index of the selected large-$R$ UFO jet.
The correlation is then computed with
\begin{equation*}
\rho=\dfrac{\sigma^{2}(\Delta_{j_{1}} + \Delta_{j_{2}})-(\sigma^{2}(\Delta_{j_{1}}) + \sigma^{2}(\Delta_{j_{2}}))}{2\sqrt{\sigma^2(\Delta_{j_{1}})}\sqrt{\sigma^2(\Delta_{j_{2}})}} \quad ,
\end{equation*}
where $\sigma^2$ represents the variance operator. The expression for $\rho$ can be derived using standard properties of the variance operator applied to a sum of distributions. The correlation values obtained from the \PYTHIA MC simulation range from 6\% at low-\pT to 2~--~3\% at high-\pT and are stable across $\eta_\text{det}$.

Taking into account the correlation term improves the closure of the asymmetry method, which can be tested by comparing the expected resolution in the MC, obtained from the balance of the \pT of UFO and particle-level jets, to the resolution on \pT obtained applying the asymmetry method to the MC simulation. The difference between the resolutions obtained in the two cases, which is taken as an additional non-closure uncertainty in the method, is reduced by up to 25\% when taking into account correlation effects.

\begin{figure}[!htp]
\centering
\subfloat[]{ \includegraphics[width=0.49\linewidth]{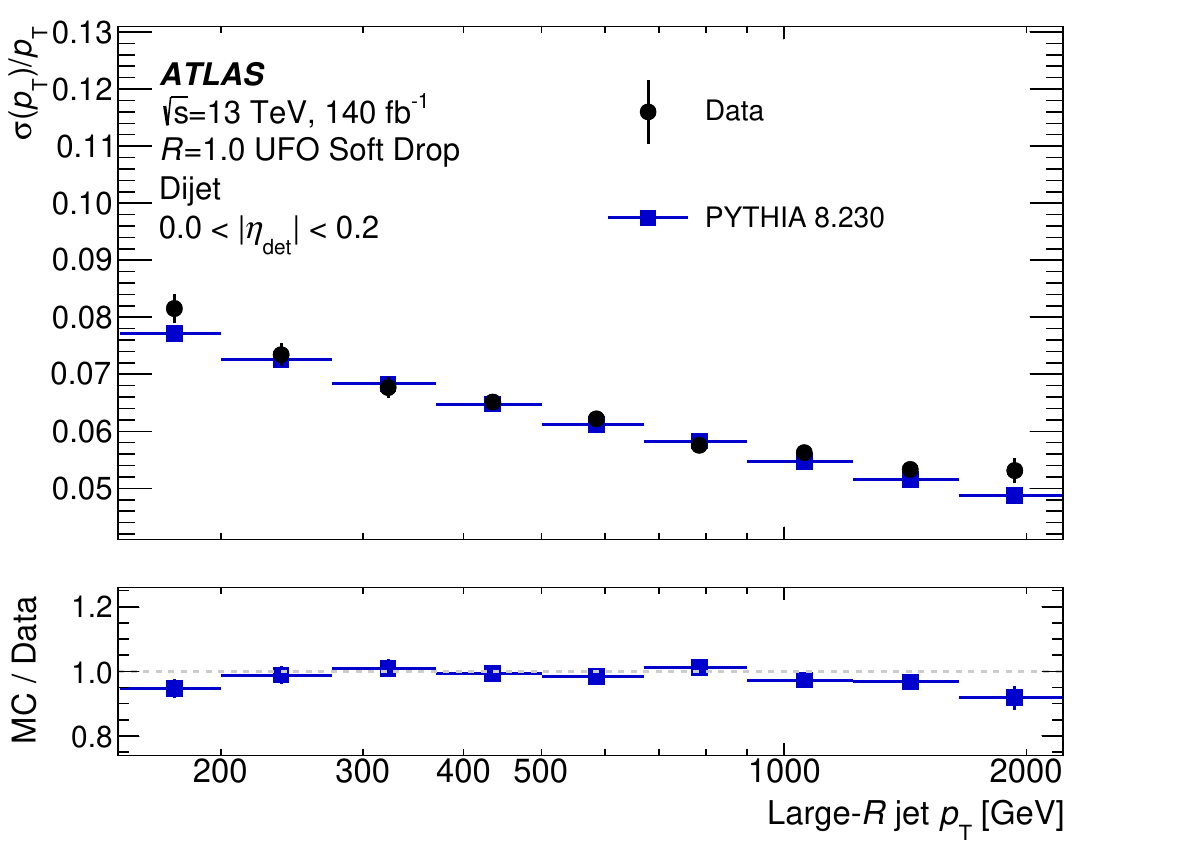}}
\subfloat[]{ \includegraphics[width=0.49\linewidth]{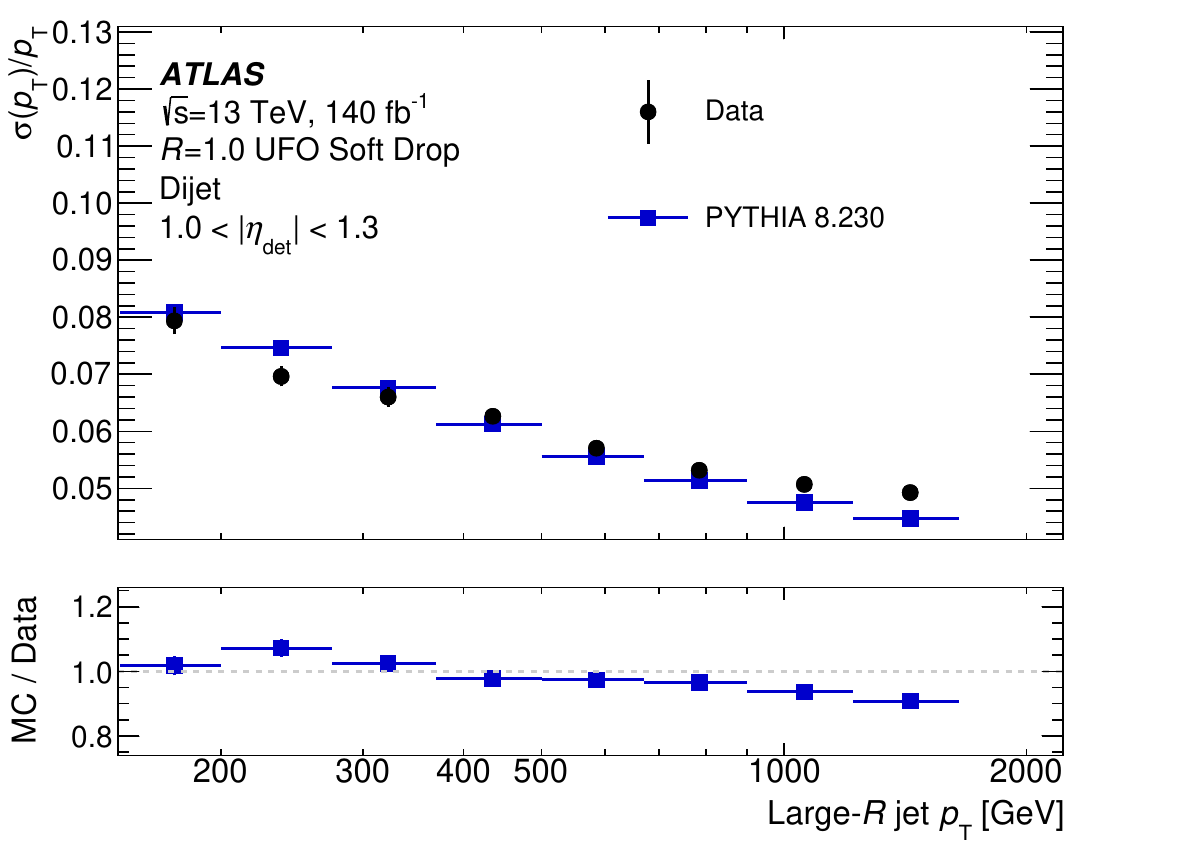}}
\caption{Large-$R$ UFO jet \pT resolution for jets with (a) $0.0<|\eta_\text{det}|<0.2$ and (b) $1.0<|\eta_\text{det}|<1.3$ as obtained from the asymmetry method in data and \PYTHIA~8 simulation.   The lower panel shows the ratio of the results obtained with Monte Carlo simulation and data. Statistical uncertainties are represented by vertical lines.}%
\label{fig:jer_res}
\end{figure}

In Figure~\ref{fig:jer_res} the relative \pt resolution extracted using the asymmetry method in data is shown. The result is compared with the relative \pt resolution as obtained with the same method from the \PYTHIA MC simulation. The resolution ranges from 8\% at 200~\gev{} to 5~--~6\% for multi-\tev{} jets in the central detector and is only degraded slightly for jets with $1.0<|\eta_\text{det}|<1.3$. The results from data agree with the ones obtained from \PYTHIA[8].

\begin{figure}[!htp]
\centering
\subfloat[]{ \includegraphics[width=0.49\linewidth]{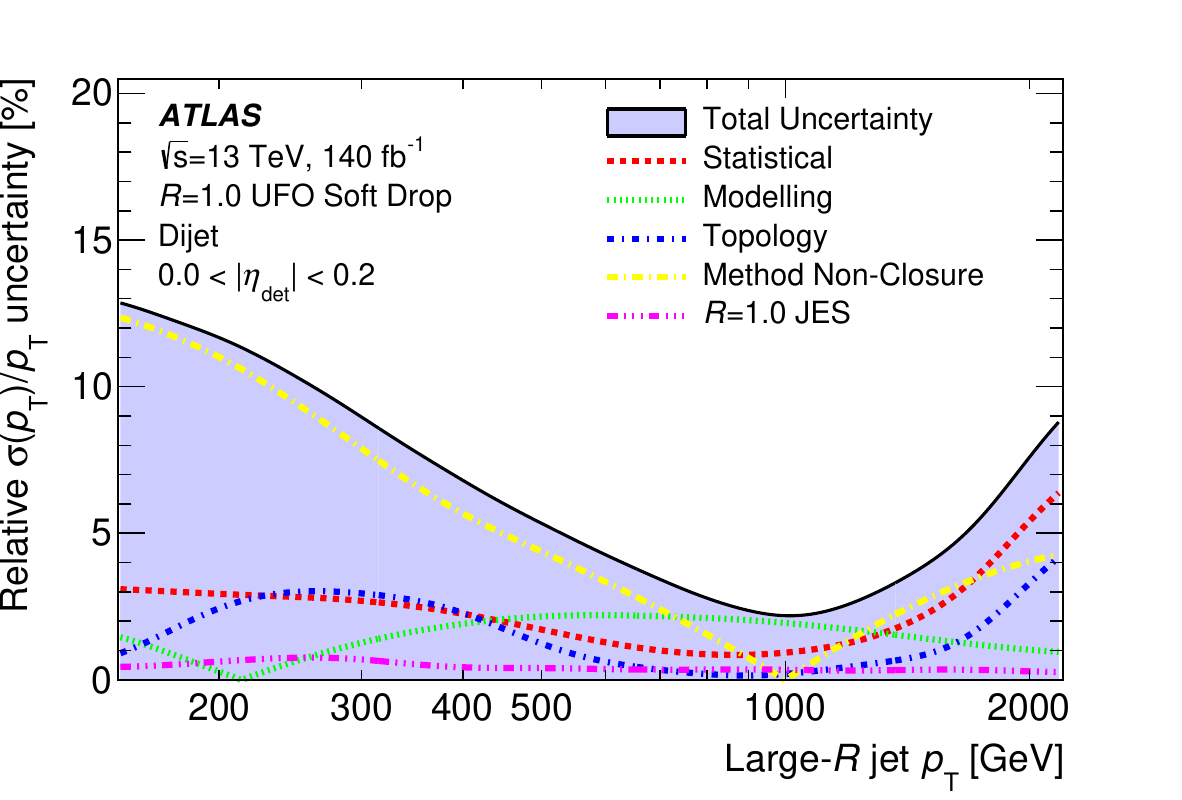}}
\subfloat[]{ \includegraphics[width=0.49\linewidth]{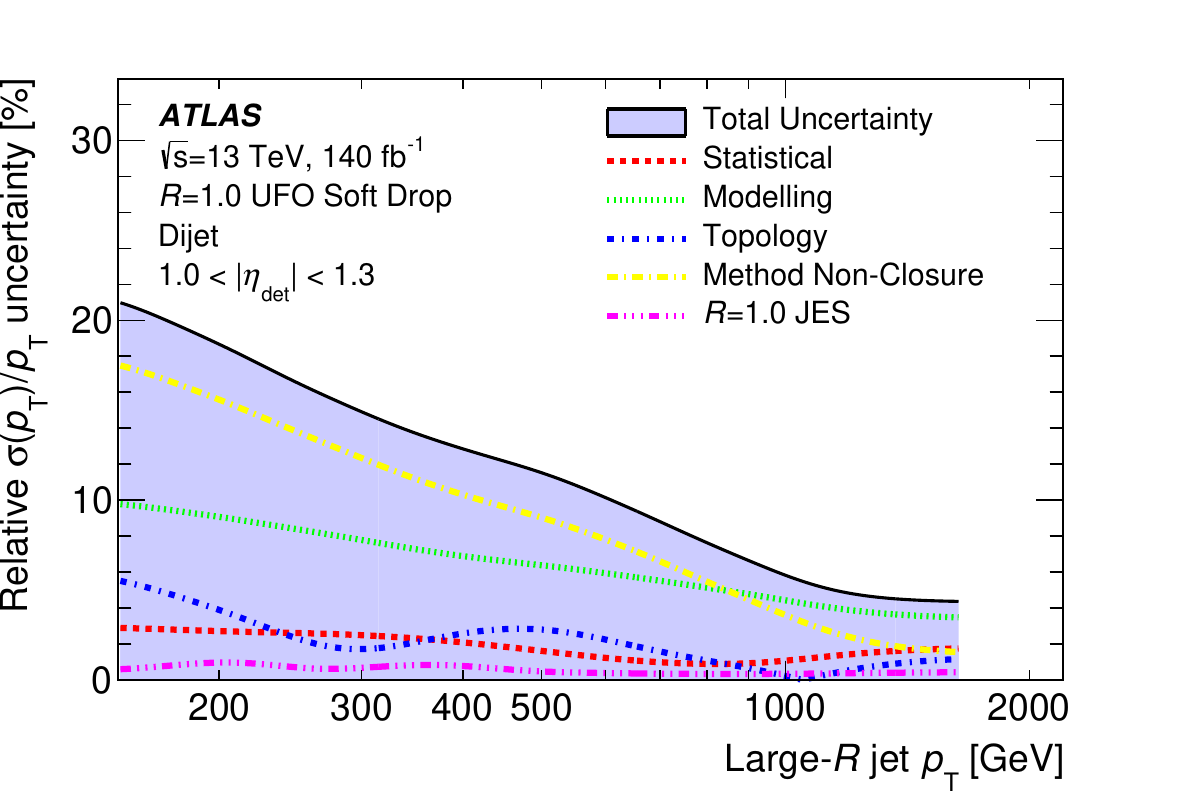}}
\caption{Breakdown of the uncertainty affecting the large-$R$ UFO jet \pT resolution measurement for (a) central  and (b) forward jets. The apparent discontinuity in the non-closure uncertainty reflects a change of sign in its definition, which is not captured when plotting the absolute uncertainty value. The $x$-axis range of the histogram in (b) is set to be the same as the histogram in (a) for consistency, even though no \pT resolution measurement is performed for forward jets with \pT around $2~\TeV$.}
\label{fig:jer_uncs}
\end{figure}

The non-closure uncertainty is the largest source of uncertainty affecting the large-$R$ UFO jet \pT relative resolution measurement, as shown in Figure~\ref{fig:jer_uncs}, with values of 10 -- 20\% around $\pt = 200~\GeV$.
Other uncertainties are related to:
\begin{itemize}
\item the cuts on $\Delta\phi$ and on the small-$R$ jet radiation (topology uncertainty) ;%
\item the generator used in the simulation, where \SHERPA[2.2.5] is used as an alternative generator and is compared with the baseline \PYTHIA[8];
\item the large-$R$ UFO JES uncertainty.
\end{itemize}

The total uncertainty in the determination of the large-$R$ UFO relative jet \pt{} resolution is shown in Figure~\ref{fig:jer_uncs} for two $|\eta_\text{det}|$ regions. The uncertainty is highest in the low-\pt{} region, and is dominated by the non-closure of the method.

The \emph{in situ} determination of the large-$R$ UFO jet relative \pt{} resolution from the momentum balance of two large-$R$ UFO jets reaches a precision of 5~--~20\%, depending on the transverse momentum and pseudorapidity. The most precise result is obtained in the central region and for transverse momenta around 1~\tev, which are the regions where the MC generator dependence of the intrinsic particle-level resolution that must be subtracted is smallest.


%
%
%

%
\section{\emph{In situ} jet mass scale calibration}
\label{sec:insitujmsr}
%

%
\label{sec:insituJMS}
The most precise calibration of the large-$R$ UFO jet mass response is achieved in a comparison of the predicted and observed peaks in the large-$R$ UFO jet mass distribution of hadronic decays of $W$ bosons and top quarks. An event topology enriched with boosted top-quark pair production is utilised for this calibration. Semileptonic \ttbar final states are considered where one top quark decays into $\mu$+jets final states to trigger and reduce backgrounds, while the other top quark decays hadronically and is reconstructed as a large-$R$ jet. Events are selected requiring an energetic muon with $\pt>25~\gev$ and $|\eta|<2.5$, small-$R$ jet  with $\pt>25~\gev$ and $|\eta|<2.5$, and a significant missing transverse momentum (above 30~\gev) from a semi-leptonically decaying top quark. The selected small-$R$ jet is required to be $b$-tagged with the DL1r algorithm~\cite{FTAG-2019-07}, at the 77\% signal efficiency working point. The leading large-$R$ UFO jet is calibrated with the factors derived in Sections~\ref{sec:mccalib} and~\ref{sec:jes_insitu}, which affect the jet mass, and is taken as the boosted-object candidate.

\begin{figure}[!ht]
\centering
\subfloat[]{\includegraphics[width=0.5\linewidth]{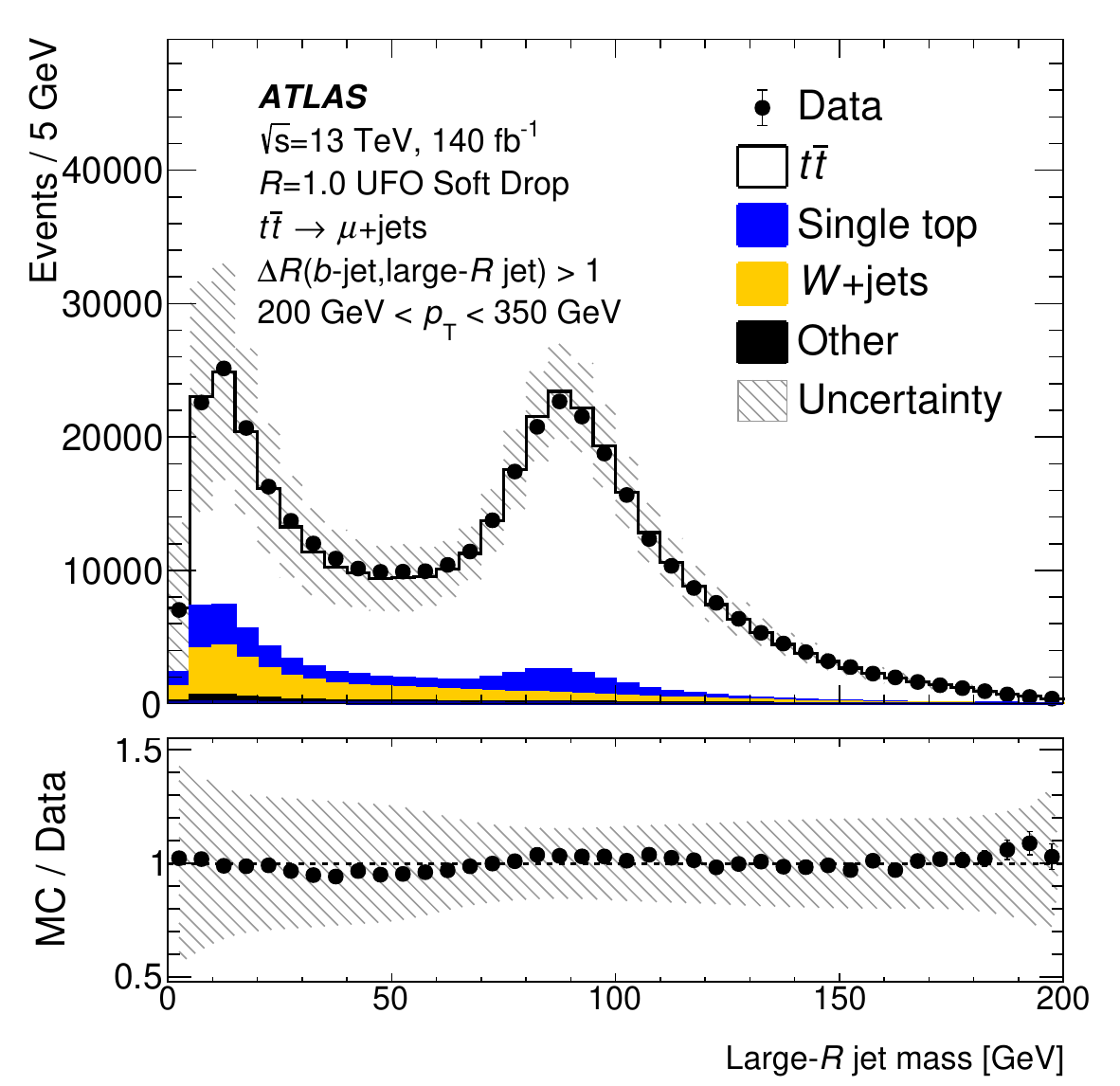}\label{fig:forward-folding-mass-peaks_a}}
\subfloat[]{\includegraphics[width=0.5\linewidth]{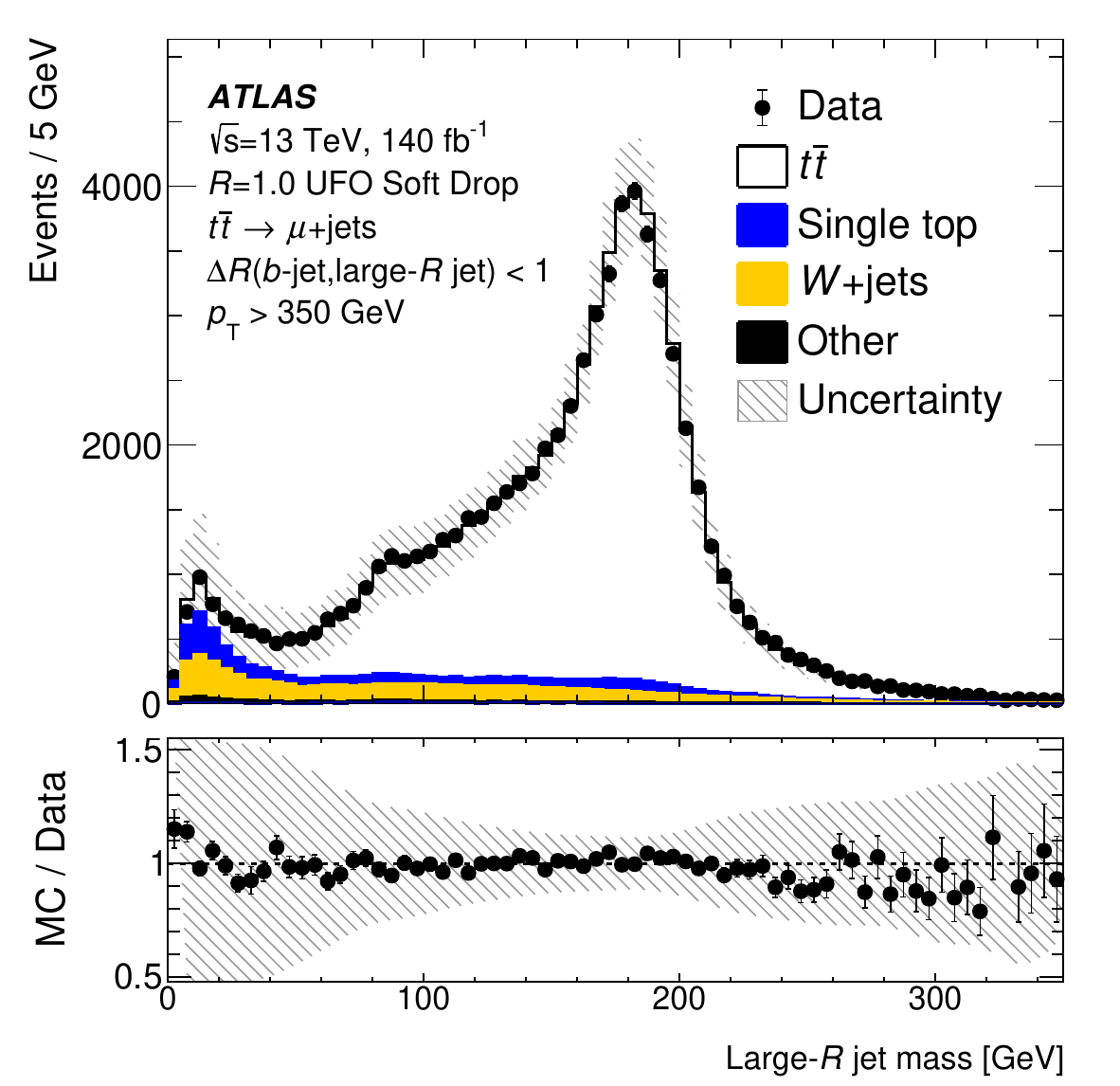}\label{fig:forward-folding-mass-peaks_b}}
\caption{The mass peaks of (a) the $W$ boson  and (b) the top quark in the large-$R$ UFO jet mass distribution of $\ttbar$ events in the $\mu+$jets channel. The observed distributions in data (solid markers) are compared with the expectation from Monte Carlo predictions for the signal and background processes. The signal sample is generated with $m_W =$ 80.4~\GeV and $m_t =$ 172.5~\GeV. The Monte Carlo expectation is normalised to the yield in data.  The lower panel shows the ratio of the results obtained with Monte Carlo simulation and data. Statistical uncertainties are represented by vertical lines, while the hashed band represents the total uncertainty in the simulation.}
\label{fig:forward-folding-mass-peaks}
\end{figure}

A clear $W$ boson mass peak is visible for large-$R$ UFO jets with relatively low \pt{} between 200~\GeV{} and 350~\GeV{}, as shown Figure~\ref{fig:forward-folding-mass-peaks_a}. In this intermediate \pt{} regime the hadronic $W$ boson decay is captured in the large-$R$ UFO jets, but the $b$-jet from the top quark decay is not contained. To further confirm this topology, events where a small-$R$ jet is tagged as a $b$-jet within $\Delta R < 1$ of the large-$R$ UFO jet are vetoed, defining a \enquote{$W$-peak} region. At higher momentum, the top quark decay is sufficiently well collimated that all decay products are captured in a single large-$R$ UFO jet. For large-$R$ UFO jets with $\pt >350~\GeV$, shown in Figure~\ref{fig:forward-folding-mass-peaks_b}, a clear top peak is observed. To enforce the fully contained topology, a $b$-tagged small-$R$ jet is required within $\Delta R < 1$ of the large-$R$ UFO jet, defining the \enquote{top-peak} region. The mass peaks are rather narrow, despite the much increased pile-up during the later stages of Run~2. Good performance is maintained through the combination of soft drop grooming and pile-up mitigation with the SoftKiller and constituent subtraction algorithms.

A quantitative estimate of the large-$R$ UFO jet mass scale and resolution is obtained from a fit to the observed distribution of MC templates with a varied large-$R$ UFO jet mass response. These templates are created by \enquote{forward folding} the particle-level jet mass distribution with a non-gaussian jet mass response based on MC simulation~\cite{JETM-2018-02,JETM-2024-05}. %
The folded jet mass is defined as
\begin{equation}
m_\text{fold} = s \, m^\text{reco} + \left( m^\text{reco} - m^\text{truth} \langle R_{m} \rangle ( m^\text{truth}, p_\text{T}^\text{truth}) \right) (r-s)
\end{equation}
where $\langle R_{m}\rangle(m^\text{truth}, p_\text{T}^\text{truth})$ represents the large-$R$ UFO jet mass response as a function of the mass and \pt of the large-$R$ particle-level jet associated with the large-$R$ UFO jet, while $m^\text{reco}$ corresponds to the large-$R$ UFO jet mass. The parameters $s$ and $r$, representing the MC-to-data relative large-$R$ UFO jet mass scale ($\text{JMS}=R^\text{m}_\text{MC}/R^\text{m}_\text{data}$) and resolution ($\text{JMR}=\sigma^\text{m}_\text{MC}/\sigma^\text{m}_\text{data}$) respectively, are varied within the ranges $0.9 < s < 1.1$ with a step size of 0.001, and $0.5 < r < 1.5$ with a step size of 0.01. The resulting templates are evaluated by a comparison to data through a $\chi^2$-fit. Assuming minimal correlation between the scale and resolution, the fit is performed by marginalizing over each parameter. The $W$-boson mass peak is fit within the range 50~\GeV{} to 120~\GeV{}, while the top peak is fit over a broader range of 120~\GeV{} to 300~\GeV{}. Despite the wider fitting range for the top peak, the impact of varying the fit range around the top-quark mass peak core was found to be small: at the per-mil level for the scale and up to a few percent for the resolution. The optimal scale and resolution values are chosen as those that minimise the $\chi^2$.

\begin{figure}[!htp]
\centering
\includegraphics[width=0.8\linewidth]{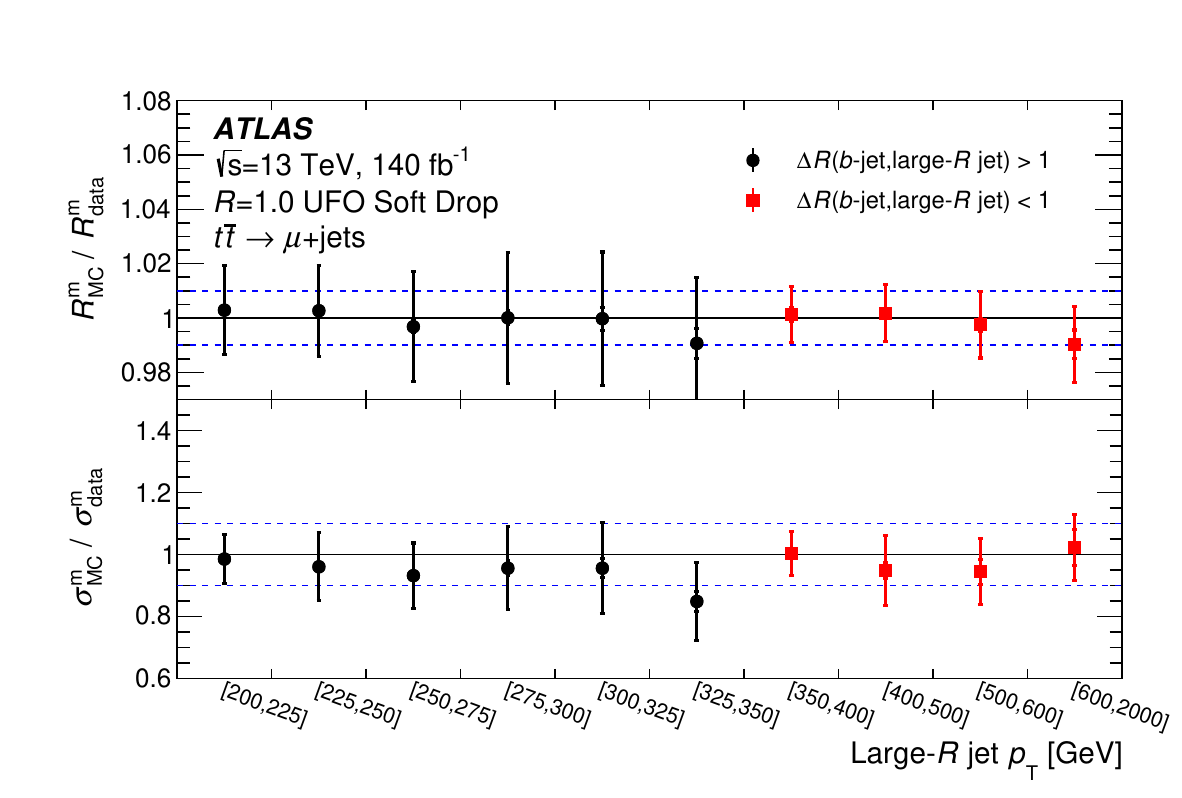}
\caption{Summary of the fit results of the jet mass scale (upper panel) and resolution (lower panel). The results from the fits in the $W$-peak region are shown for $\pT<350~\GeV$, the fits in the top peak region for $\pT>350~\GeV$. The error bars represent the total uncertainty, which is dominated by systematic uncertainties in the modelling of the \ttbar{} signal. Where visible, statistical uncertainties are indicated with a second, smaller error bar. The measurement is performed for jets that have their four-momentum corrected by the \emph{in situ} momentum calibration.}
\label{fig:forward-folding-summary}
\end{figure}

The simultaneous fits to JMS and JMR are performed in six bins of large-$R$ UFO jet \pt for the $W$-peak region and in four \pt bins for the top peak region, as shown in Figure~\ref{fig:forward-folding-summary}.
The total uncertainty is primarily driven by systematic uncertainties, which are largely dominated by the difference in parton shower modelling between \POWPY[8] and \POWHER[7]. The uncertainty on the value of the top-quark mass is evaluated by a variation of $\pm 0.5~\GeV$  from the nominal value in the MC simulation of $m_t = 172.5 \GeV$.
The effect of the \emph{in situ} JES uncertainty on the mass response measurement was found to be small, few per-mille for the JMS and around 2\% for the JMR.
The JMS results are compatible with unity well within the total uncertainty, indicating that the MC simulation provides a good description of the jet mass response, after the {\em in situ} correction derived in Section~\ref{sec:jes_insitu} is applied to the jet four-vector. The MC simulation also provides an adequate description of the jet mass resolution.

\begin{figure}[!htp]
\centering
\subfloat[]{\includegraphics[width=0.49\linewidth]{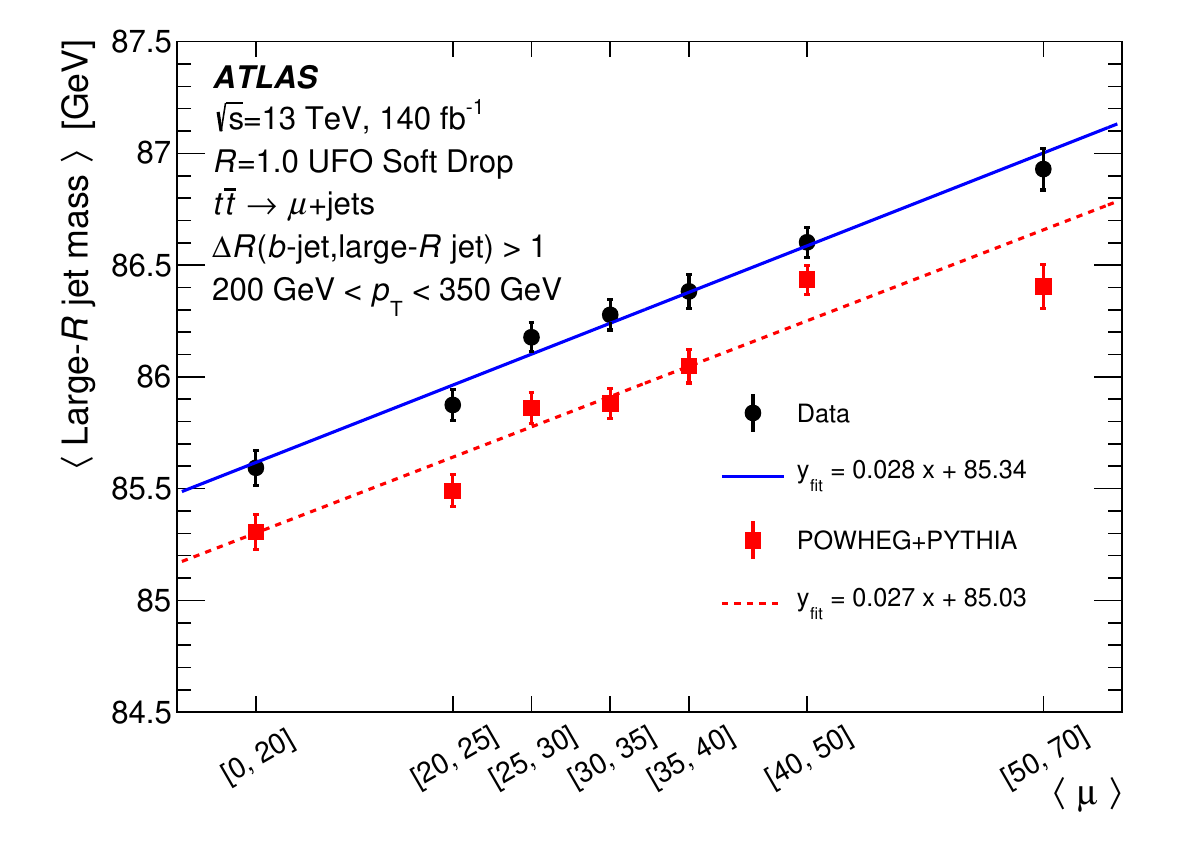}\label{fig:forward-folding-pile-up_a}}
\subfloat[]{\includegraphics[width=0.49\linewidth]{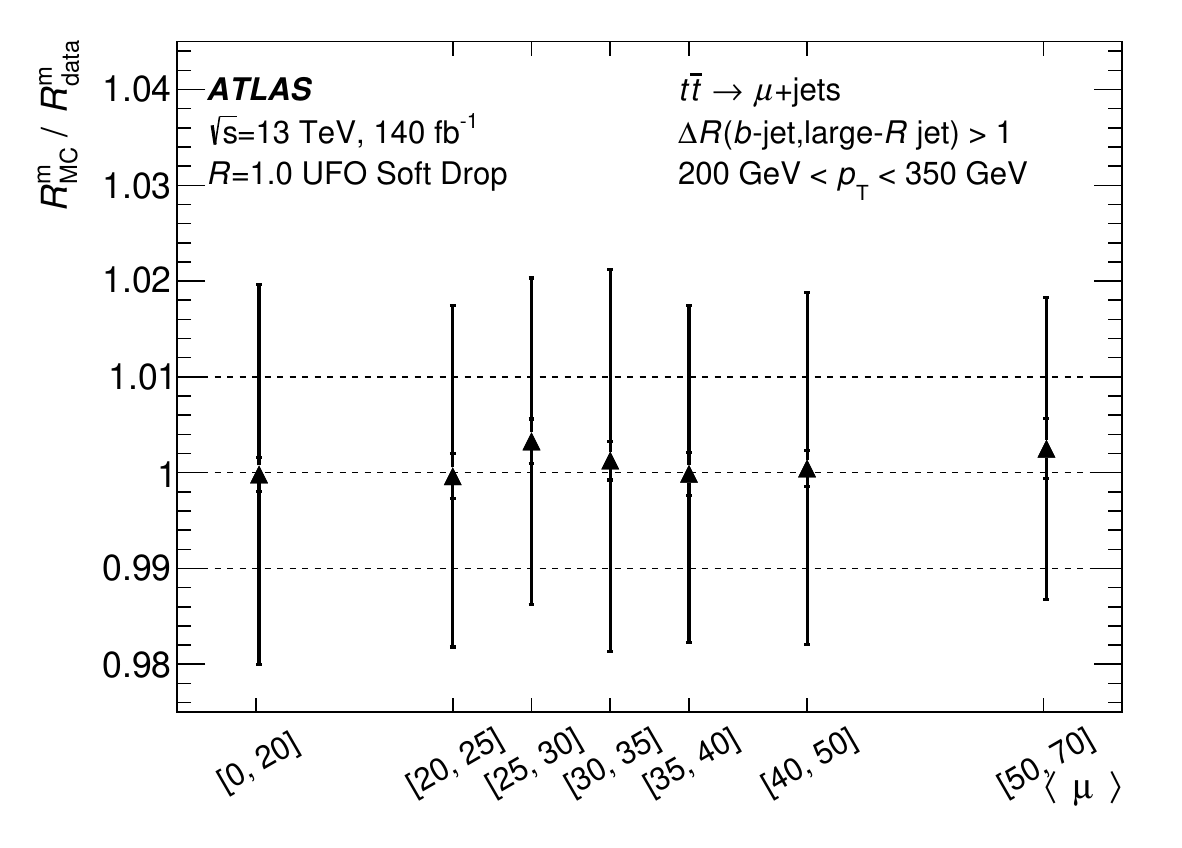}\label{fig:forward-folding-pile-up_b}}
\caption{(a) The average large-$R$ UFO jet mass as a function of the average number of $pp$ interactions, for the $W$-peak region. The impact of pile-up in experimental data is compared with that in Monte Carlo simulations, by a linear fit to each of the distributions. (b) The large-$R$ UFO jet mass scale extracted via the forward folding method for the $W$-peak region, as a function of the average number of pile-up interactions $\langle \mu \rangle$ per bunch crossing. %
The small error bars represent the statistical uncertainty while the large error bars show the total uncertainty.
}
\label{fig:forward-folding-pile-up}
\end{figure}

The pile-up resilience of the large-$R$ UFO jet mass response is tested by dividing the sample into seven subsamples with an average number of pile-up interactions $\langle\mu\rangle$ ranging from 0~to~70, a range that covers the values of $\langle\mu\rangle$ produced by the LHC during Run~2. The average jet mass for the large-$R$ UFO jets in the $W$-peak region with mass interval from 50~\GeV{} to 120~\GeV{} has a small pile-up dependence, as shown in Figure~\ref{fig:forward-folding-pile-up_a}: an increase in the $W$-boson peak position of approximately 30~\MeV per additional pile-up interaction is observed. The fits show a compatible pile-up dependence in data and simulations, indicating that pile-up effects are modelled well in the MC predictions.
The forward folding method is applied to each of the seven $\langle \mu\rangle$ subsamples to measure a potential $\langle\mu\rangle$ dependence of the data-to-MC relative large-$R$ UFO jet mass response, in the $W$-peak region. Figure~\ref{fig:forward-folding-pile-up_b} displays the large-$R$ UFO jet mass scale values obtained in each bin for samples divided in $\langle \mu\rangle$, suggesting that the observed pile-up effect in data and its modelling in MC simulations have a negligible impact on the resulting jet mass scale.%

\FloatBarrier


\section{Combination and uncertainties}
\label{sec:comb_and_uncs}
%


The measurements of the large-$R$ UFO jet \pT and mass response are performed in bins of the jet transverse momentum (the $\pT^\text{ref}$ are translated to the large-$R$ UFO jet \pT) and evaluated inclusively in mass. %
Each of the measurements is limited in \pT reach, as they depend on the final state in which the measurement is performed.
Different measurements of the large-$R$ UFO  jet response can be combined together with a statistical analysis, following the methodology already employed in Refs.~\cite{JETM-2018-05} and~\cite{PERF-2012-01}, to provide a continuous calibration curve covering the largest possible phase space.

The combination method proceeds in
steps that take into account correlations between uncertainties between the \emph{in situ} methods.
First, the relative \pT or mass response is defined in fine \pT bins, separately for each \emph{in situ} method, using interpolating splines based on first- or second-order polynomials.
Secondly, the actual combination is carried out using a weighted average of the \emph{in situ} measurements based on a $\chi^2$-minimisation. The $\chi^2$ test is also useful to define the level of agreement between \emph{in situ} measurements where they overlap. The weights in the average take into account the statistical and systematic uncertainties, as well as correlations and differing bin sizes.
Pseudo-experiments are employed to consistently propagate all uncertainties into the evaluation of the average. They are created to represent the ensemble of measurements, including the interpolation and averaging step. They are generated taking into account all known correlations by coherently shifting all correction factors by one standard deviation. The difference between the shifted-correction result and the nominal result provides an estimate of the propagated systematic uncertainty. Each source of uncertainty is taken as fully correlated across kinematic regions (i.e. as a function of \pT and $\eta$) and uncorrelated with other sources.
A smoothing procedure using a variable-size sliding interval with a gaussian kernel is applied to the combined response ratio and its associated systematic uncertainties. This smoothing removes spikes due to statistical fluctuations in the measurements, as well as discontinuities at the first and last point in a given measurement.

\subsection{JES combination}
\label{sec:jes_combination}

\begin{figure}[!htp]
\centering
\includegraphics[width=0.9\linewidth]{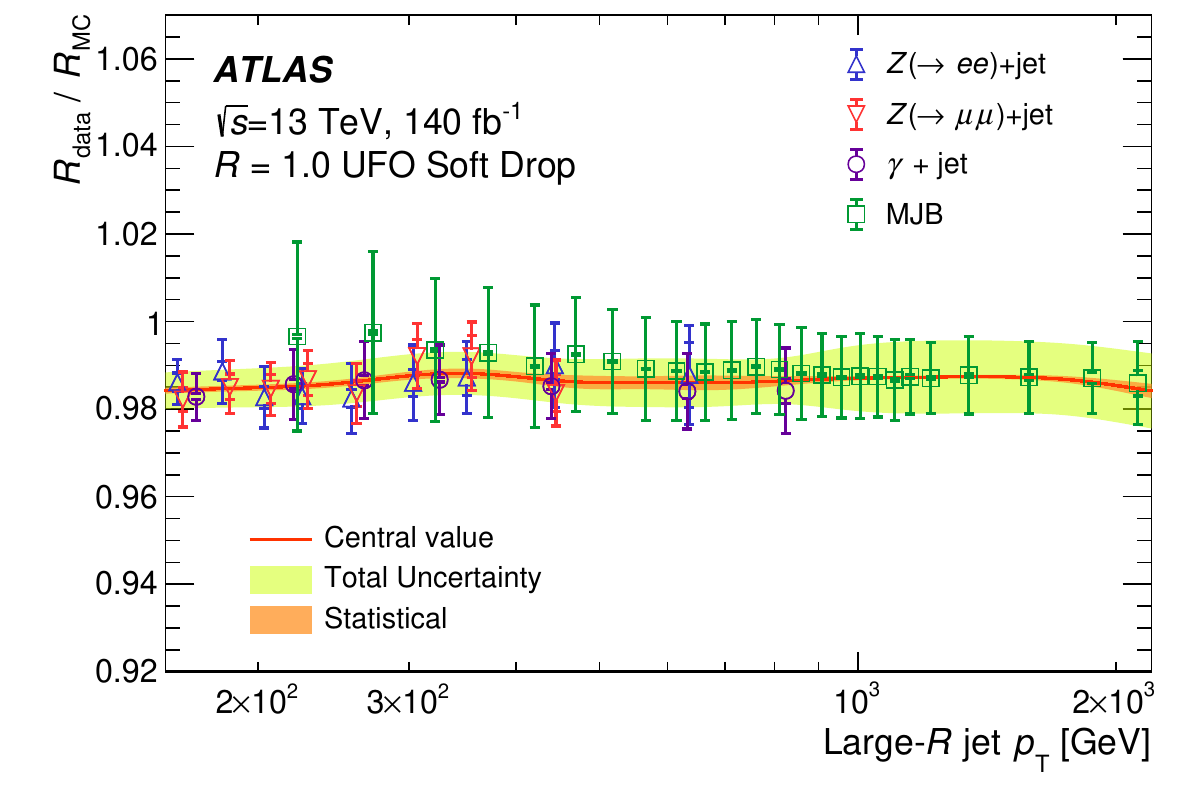}
\caption{Results of the {\em in situ} JES balance measurements. The markers correspond to the results of the $Z+$jet method (up-pointing markers for the $Z\rightarrow ee$ channel, down-pointing markers for the $Z\rightarrow \mu\mu$ one), the $\gamma +$jet method (round markers) and the multijet balance method (squared markers). The error bars represent the total uncertainty in each measurement, with the inner ticks on the bars representing the statistical uncertainty only. The central value of the {\em in situ} correction is indicated with a continuous line, while its statistical and total uncertainties are represented by the filled bands around it.}
\label{fig:jes_combination_data_points}
\end{figure}

The results of the \emph{in situ} JES calibration methods described in Section~\ref{sec:zjets} through~\ref{sec:multijetbalance} are combined. The JES response is measured in various final states covering a  150 -- 2400~\GeV range in large-$R$ UFO  jet \pT.  The very precise results from the $Z+$ jet method for large-$R$ UFO jet \pt{} below about 500~\gev{} are complemented by results from $\gamma +$ jet events up to 900~\GeV, and the results of the multijet balance that extend up to $\pt \sim 2.4~\tev$.
The results of the three methods are shown in Figure~\ref{fig:jes_combination_data_points} together with the combination. Good compatibility is observed with an average value of $R_{\mathrm{data}}/R_{\mathrm{MC}} \sim0.985$ and no significant \pt dependence.
The JES combination result is used to calibrate the four-momentum of large-$R$ UFO jets in data.

\begin{figure}[!htp]
\centering
\subfloat[]{\includegraphics[width=0.49\linewidth]{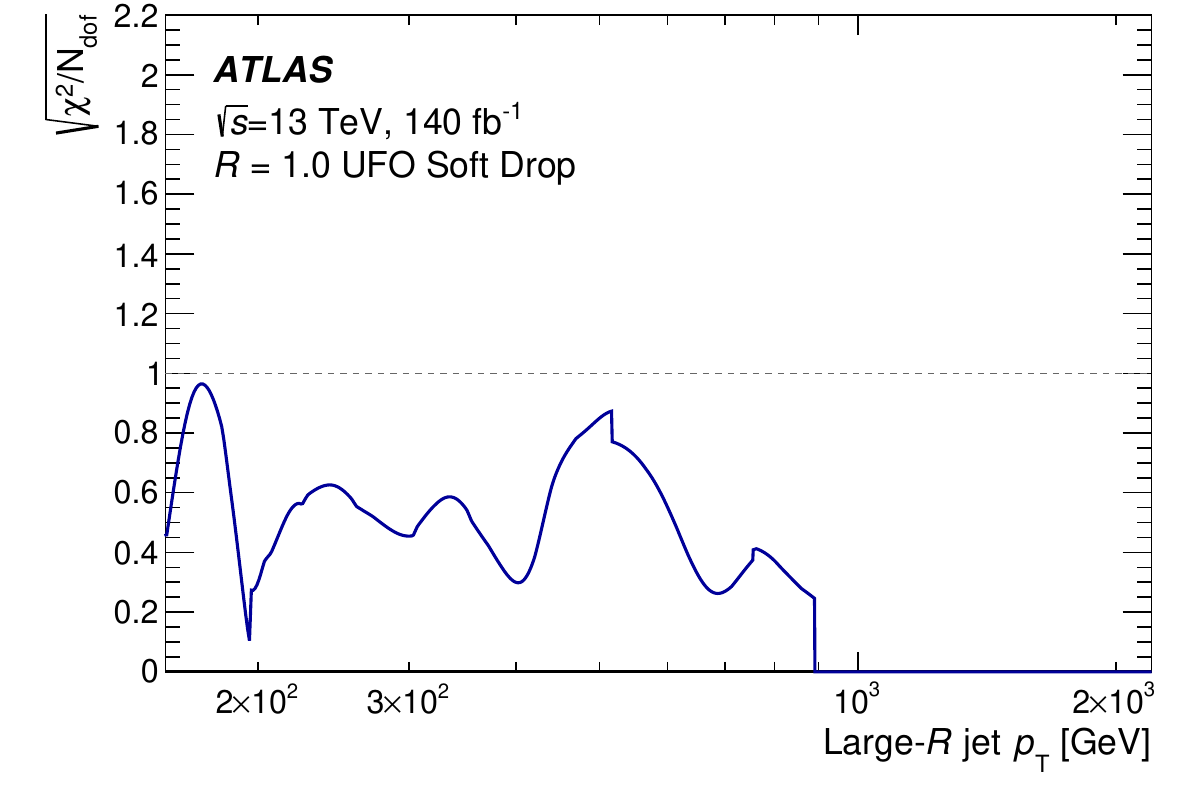}\label{fig:jes_combination_chi2_weights_a}}
\subfloat[]{\includegraphics[width=0.49\linewidth]{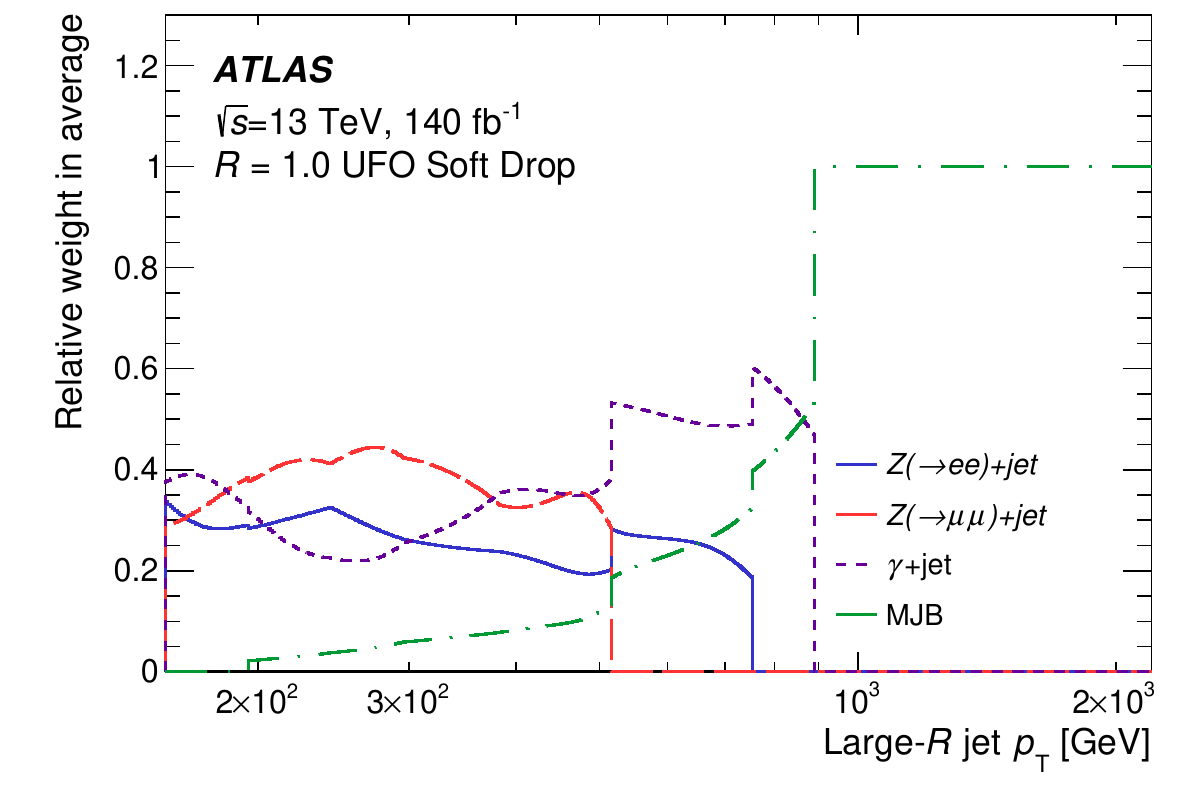}\label{fig:jes_combination_chi2_weights_b}}
\caption{The (a) $\chi^2$ per degree of freedom value and (b)  weights returned by the combination of the results of the balance methods. }
\label{fig:jes_combination_chi2_weights}
\end{figure}

Whenever multiple methods are available in a given \pt{} bin, the compatibility of the results can be expressed quantitatively.
The $\chi^2$ per degree of freedom ($\chi^2/N_\text{dof}$)  value of the combination, shown in Figure~\ref{fig:jes_combination_chi2_weights_a} confirms the excellent agreement of the four methods. Figure~\ref{fig:jes_combination_chi2_weights_b} shows the weights assigned to the four methods. The  $R_{Z+\text{jets}}$ measurements, in the $Z\rightarrow e^+e^-$ and $Z\rightarrow \mu^-\mu^+$  channels separately, have similar weights as the $R_{\gamma+\text{jets}}$ measurements for large-$R$ UFO jet \pt{} below 500~\gev{}. At higher \pt{}, the $R_{\gamma+\text{jets}}$ measurements dominate the combination, until the $R_{\text{MJB}}$ results take over at the highest momenta, above 1~\tev.

\FloatBarrier

\subsection{JMS extrapolation}
\label{sec:jms_combination}

The large-$R$ UFO jet mass scale uncertainties are extrapolated to a higher \pt than the one covered in Section~\ref{sec:insituJMS}, taking advantage of previous results on large-$R$ calorimeter jets from Ref.~\cite{JETM-2018-02} based on the $R_\text{trk}$ method in multijet events.

The JMS measurement with the $R_\text{trk}$ method relies on the fact that the jet mass and the mass computed from tracks associated with the jet are two independent quantities. This does not apply to large-$R$ UFO jets, since the tracks are used in the definition of the jet constituents. Hence, it is not possible to directly apply the method to large-$R$ UFO jets.
On the other end, the $R_\text{trk}$ measurements can be performed in multijet events that cover a complementary phase space to the one selected in Section~\ref{sec:insituJMS}.
An additional uncertainty is added in order to use the $R_\text{trk}$ results based on calorimeter jets as a JMS measurement on large-$R$ UFO jets, to cover differences arising from the choice of inputs and algorithms used in the jet definition of large-$R$ UFO jets and calorimeter jets.
This uncertainty is obtained by comparing the jet mass response for the two different jet definitions in MC simulation. The comparison is performed differentially in the associated truth jet \pT and mass. The uncertainty computed this way is found to be up to $\approx 10\%$ for $m<50~\GeV$, and $\leq 1\%$ otherwise.

The additional uncertainty allows for extrapolating the large-$R$ UFO jet mass scale to regions where there are no precision constraints: at high \pt{} and for jet masses below and above the intervals covered by the $W$ boson and top quark mass peaks.
In the mass range overlapping with the one covered by the large-$R$ UFO jet mass scale measurements of Section~\ref{sec:insituJMS}, the $R_\text{trk}$ large-$R$ calorimeter jet measurements are combined to the large-$R$ UFO jet ones obtained with forward folding, with a procedure similar to that used for the jet energy scale of Section~\ref{sec:jes_combination}.
The results from the forward folding on large-$R$ UFO jets and $R_\text{trk}$ on large-$R$ calorimeter jets agree, yielding a $\chi^2$ per degree of freedom well below unity in the combination. Where available, the fits to $W$-boson and top-quark mass peaks dominate the combination.

\begin{figure}[!htp]
\centering
\includegraphics[width=0.75\linewidth]{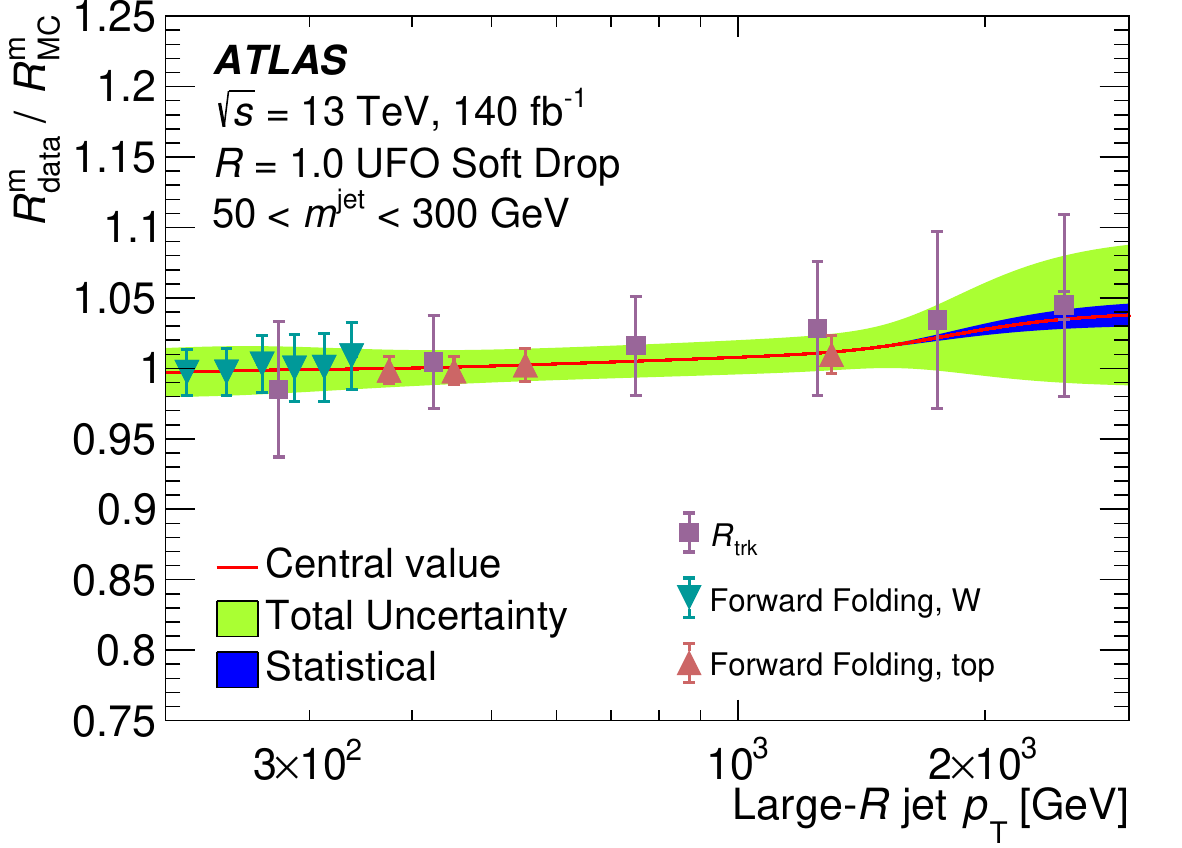}
\caption{Combination of \emph{in situ} measurements of the jet mass scale for jet masses between 50~\GeV{} and 300~\GeV{}, and as a function of \largeR{} jet \pt{}. The jet mass scale measurements are obtained from jets on which the \emph{in situ} jet energy scale correction is already applied. Triangle-shaped markers represent the measurements based on forward folding fits to the $W$-boson mass peak (down-pointing markers) and the top quark mass peak (up-pointing markers). The square markers represent the results obtained with the $R_\text{trk}$ method and extrapolate the jet mass scale to high \pt{} and jet masses below the $W$-boson mass and above the top quark mass. The error bars represent the total uncertainty in each measurement, with the inner ticks on the bars representing the statistical uncertainty only. The continuous line shows the combination of all results, and the filled bands around it represent the total and statistical uncertainty in the combination. }
\label{fig:jmscombination_jms_vs_pt}
\end{figure}

Figure~\ref{fig:jmscombination_jms_vs_pt} shows the result of the jet mass scale combination in the jet mass interval between 50~\GeV{} and 300~\GeV{}, where precision \emph{in situ} measurements are available from the forward folding fits to the $W$-boson and top-quark mass peaks.
The JMS combination is compatible with unity, well within the uncertainty, after the application of the \emph{in situ} JES correction derived in Section~\ref{sec:jes_combination}. As a consequence, no further \emph{in situ} correction is applied to the large-$R$ UFO jet mass.

\FloatBarrier

\FloatBarrier

\subsection{Topology and flavour uncertainty}
\label{subsec:topology}
The \emph{in situ} measurements of the large-$R$ UFO jet energy and mass response are performed on samples dominated by definite jet topologies and flavours. The balance methods used to measure the JES are primarily sensitive to light quarks, while the forward folding analysis determines the mass response on boosted $W$ bosons and top quarks.
Additional uncertainties are assigned to the JES and JMS to cover for possible differences between the detector response for large-$R$ jets from different sources and of different flavours.

The uncertainty associated to flavour composition is introduced to account for the difference in jet energy response of quark- and gluon-initiated jets in a sample with some uncertainty in the fraction of gluon-initiated jets.
The uncertainty associated to flavour response  accounts for the fact that, unlike the response to quark-initiated jets, the response to gluon-initiated jet is found to differ significantly between generators. Both of these uncertainties affect the JES, and are parameterised in jet $\eta$ and \pt and evaluated following the strategy developed in Ref.~\cite{JETM-2018-05}, using  MC simulations with \PYTHIA[8] and the angular-ordered parton shower in \HERWIG[7]. For jets in a sample where the flavour composition is unknown, the JES uncertainty due to flavour composition is found to range from 0.6\% to 0.8\%, while the JES uncertainty due to flavour response ranges from 0~to 0.5\%.

The topology uncertainty accounts for different splitting patterns and energy distributions within the jets, and it affects both the jet energy and jet mass responses.
The jet energy topology uncertainty is parameterised as a function of the jet $\eta$ and \pt and is applied to jets originating from a heavy resonance, since all the energy response measurements are carried out on samples dominated by light-quark and gluon jets. The jet energy topology uncertainty is estimated by taking the absolute difference in energy response of jets originating from a top-quark or a $W$ boson between two MC simulations. The \POWPY[8] and \POWHER[7] \ttbar simulations are used to evaluate the uncertainty associated to jets from top quarks, while the $W(\rightarrow qq)\gamma$ samples generated with \SHERPA and \MGPY[8] are used to evaluate the uncertainty on jets from $W/Z$ bosons.
T The uncertainty in the jet energy scale ranges from 0~to 1.5\%, with the higher values for the phase space where the jet \pt is around 1~\TeV, while the uncertainty in the energy resolution is found to be negligible.
The jet mass topology uncertainty affects large-$R$ jets stemming from quarks or gluons, as the \emph{in situ} measurement is carried out on jets that contain the decay products of boosted top-quarks and $W$ bosons. The jet mass topology uncertainty is parameterised against the jet mass and \pt, which are the same variables considered in the JMS uncertainty extrapolation, and is estimated by taking the absolute difference in mass response from jets in the \PYTHIA[8] and the angular-ordered \HERWIG[7] multijet MC samples.
The topology uncertainty for the JMS response ranges from 1\% for jets with high mass and \pt, to 4\% for jets with low \pt. The topology uncertainty for the JMR response is found to be up to 20\% for large-$R$ UFO jets with mass below $100~\GeV$ and 1\% for masses above 300~\GeV.

\FloatBarrier

\subsection{Nuisance parameter reduction}
\label{subsec:reduction}

The combination of different \emph{in situ} methods described in Section~\ref{sec:jes_combination} yields a precise JES determination, with nearly two hundred sources of uncertainty. To keep the propagation of the JES uncertainty to ATLAS analyses manageable, the number of nuisance parameters is reduced. This procedure aims to maintain a good representation of the correlations of the systematic uncertainties on the \emph{in situ} measurements. A set of independent nuisance parameters describing the main uncertainty components is obtained following the method described for the first time in Ref.~\cite{PERF-2014-02}.
The total covariance matrix of the JES calibration, which includes all the \emph{in situ} uncertainty sources, is diagonalized and a new set of independent uncertainty sources is derived from the eigenvectors and eigenvalues.
Only uncertainties depending on a single parameter (the jet \pt) are combined and any uncertainties with dependencies on other parameters, such as the ones affecting the $\eta$-intercalibration measurement, are left separate.
In total, 194 uncertainty sources are used as inputs to the reduction procedure.
To preserve knowledge of the uncertainty source in this procedure, the reduction is done within categories: statistical, detector,
modelling and mixed. This facilitates the correlation between experiments and different data-taking years to be assessed in combinations of measurements. A new uncertainty scheme labelled as \enquote{category reduction} is defined, and includes 23 uncertainty sources split into the previously mentioned categories.
The difference in jet-to-jet correlations induced by the full uncertainty model and the category reduction one is kept below 1\%.
%

\FloatBarrier


\section{Conclusions}
\label{sec:conclusions}
%


This paper presents the ATLAS calibration chain for large-radius jets in Run~2 of the LHC\@. These jets are reconstructed with the anti-$k_t$ algorithm with radius parameter $R=1.0$ from Unified Flow Objects that combine the tracker and calorimeter information. The soft drop grooming together with the constituent subtraction and SoftKiller algorithms are applied to mitigate the impact of pile-up and soft radiation.

The first calibration step corrects the reconstructed jet energy, pseudorapidity and invariant mass to the particle level equivalents. Excellent closure is found for the jet energy, while the mass response generally closes within 10\%, with a modest residual dependence on jet topology. The expected jet mass resolution reaches as low as 10\% for jet masses of 100 -- 200~\gev{} and \pt $< $ 2~\tev, improving significantly over the resolution of purely calorimeter-based jets.

The second step of the calibration corrects for  differences between data and simulation arising from imperfect modelling of the detector response. Dijet events in different regions of the detector are used to derive the $\eta$-intercalibration that ensures a uniform response as a function of pseudorapidity. Several processes where a jet recoils against a well-measured reference system (a $Z$ boson, a photon or several lower-\pt{} jets) form the heart of the \emph{in situ} calibration, which corrects the \pt{} response for differences between data and MC predictions in the central region of the detector. Combining their results, the jet energy scale is determined \emph{in situ} with a precision better than 1\% over a broad \pt{} range from 175~\gev{} to 1~\tev{}. For large-radius jets with transverse momenta up to 2~\tev{} the uncertainty in jet energy response reaches the $\leq2\%$ level.

The jet energy resolution is also validated using collision data, taking advantage of the momentum balance of two jets in multijet events. The relative transverse momentum resolution $\sigma ( \pt) / \pt $ in the central part of the detector ranges from about 8\% at 150~\gev{} to 5\% for energetic jets with \pt{} between 1~and 2.2~\tev. The jet energy resolution predicted by the MC model is confirmed within 10 -- 20\% uncertainty.

The jet mass response is calibrated with fits to the jet mass distribution in samples with hadronic decays of boosted $W$ bosons and top quarks. In the \pt{} and mass range amenable to this technique, a precision of about 1\% is achieved, limited by modelling uncertainties.

The results in this paper enable a host of searches and measurements with large-radius jets. The precision of the calibration has improved significantly in key areas relative to the first calibration with partial Run~2 data. Such improvements were driven by the largest data sample considered, the usage of a more advanced definition for large-$R$ jets, progress in the modelling of various reference processes, and improvements in the calibration techniques used.
The results in this paper also form the basis of the large-radius calibrations that are used in ATLAS analyses using the Run~3 data.


%
\section*{Acknowledgements}
%

%
%

%
%

We thank CERN for the very successful operation of the LHC and its injectors, as well as the support staff at
CERN and at our institutions worldwide without whom ATLAS could not be operated efficiently.

The crucial computing support from all WLCG partners is acknowledged gratefully, in particular from CERN, the ATLAS Tier-1 facilities at TRIUMF/SFU (Canada), NDGF (Denmark, Norway, Sweden), CC-IN2P3 (France), KIT/GridKA (Germany), INFN-CNAF (Italy), NL-T1 (Netherlands), PIC (Spain), RAL (UK) and BNL (USA), the Tier-2 facilities worldwide and large non-WLCG resource providers. Major contributors of computing resources are listed in Ref.~\cite{ATL-SOFT-PUB-2026-001}.

We gratefully acknowledge the support of ANPCyT, Argentina; YerPhI, Armenia; ARC, Australia; BMWFW and FWF, Austria; ANAS, Azerbaijan; CNPq and FAPESP, Brazil; NSERC, NRC and CFI, Canada; CERN; ANID, Chile; CAS, MOST, NSFC and FRFCU, China; Minciencias, Colombia; MEYS CR, Czech Republic; DNRF and DNSRC, Denmark; IN2P3-CNRS and CEA-DRF/IRFU, France; SRNSFG, Georgia; BMFTR, HGF and MPG, Germany; GSRI, Greece; RGC and Hong Kong SAR, China; ICHEP and Academy of Sciences and Humanities, Israel; INFN, Italy; MEXT and JSPS, Japan; CNRST, Morocco; NWO, Netherlands; RCN, Norway; MNiSW, Poland; FCT, Portugal; MNE/IFA, Romania; MSTDI, Serbia; MSSR, Slovakia; ARIS and MVZI, Slovenia; DSI/NRF, South Africa; MICIU/AEI, Spain; SRC and Wallenberg Foundation, Sweden; SERI, SNSF and Cantons of Bern and Geneva, Switzerland; NSTC, Taipei; TENMAK, T\"urkiye; STFC/UKRI, United Kingdom; DOE and NSF, United States of America.

Individual groups and members have received support from BCKDF, CANARIE, CRC and DRAC, Canada; CERN-CZ, FORTE and PRIMUS, Czech Republic; COST, ERC, ERDF, Horizon 2020 and Marie Sk{\l}odowska-Curie Actions, European Union; Investissements d'Avenir Labex, Investissements d'Avenir Idex and ANR, France; DFG and AvH Foundation, Germany; Herakleitos, Thales and Aristeia programmes co-financed by EU-ESF and the Greek NSRF, Greece; BSF-NSF and MINERVA, Israel; NCN and NAWA, Poland; La Caixa Banking Foundation, CERCA and AGAUR programs from Generalitat de Catalunya and PROMETEO and GenT Programmes Generalitat Valenciana, Spain; G\"{o}ran Gustafssons Stiftelse, Sweden; The Royal Society and Leverhulme Trust, United Kingdom; Eric and Wendy Schmidt Fund for Strategic Innovation, United States of America.

In addition, individual members wish to acknowledge support from Chile: Agencia Nacional de Investigaci\'on y Desarrollo (ANID FONDECYT reg. 1230987, FONDECYT 1230812, FONDECYT 1240864, Fondecyt 3240661, Fondecyt Regular 1240721); China: Fundamental Research Funds for the Central Universities (010-63263105), Chinese Ministry of Science and Technology (MOST-2023YFA1605700, MOST-2023YFA1609300), National Natural Science Foundation of China (NSFC 12275265, NSFC-W2543005); Czech Republic: Czech Science Foundation (GACR - 24-11373S), Ministry of Education Youth and Sports (ERC-CZ-LL2327, FORTE CZ.02.01.01/00/22\_008/0004632); EU: H2020 European Research Council (ERC - 101002463); European Union: European Research Council (BARD No. 101116429, ERC - 101219398, ERC 101089007), European Regional Development Fund (HE COFUND GA No.101081355, ERDF), Marie Sklodowska-Curie Actions (GAP-101168829); France: Agence Nationale de la Recherche (ANR-22-EDIR-0002, ANR-24-CE31-0504-01); Germany: Deutsche Forschungsgemeinschaft (DFG - 469666862); China: Research Grants Council (GRF); Italy: Istituto Nazionale di Fisica Nucleare (LHC-MIUR - 28003/2025), Ministero dell'Università e della Ricerca (NextGenEU I53D23000820006 M4C2.1.1, SOE2024\_0000023); Japan: Japan Society for the Promotion of Science (JSPS KAKENHI  JP25H0063, JSPS KAKENHI JP22H04944, JSPS KAKENHI JP24K23939, JSPS KAKENHI JP24KK0251, JSPS KAKENHI JP25H00650, JSPS KAKENHI JP25H01291, JSPS KAKENHI JP25K01011, JSPS KAKENHI JP25K01023, JSPS KAKENHI JP25KK0047); Poland: Polish National Science Centre (NCN 2021/42/E/ST2/00350, NCN OPUS 2023/51/B/ST2/02507, NCN OPUS nr 2022/47/B/ST2/03059, NCN UMO-2019/34/E/ST2/00393, UMO-2022/47/O/ST2/00148, UMO-2023/49/B/ST2/04085, UMO-2023/51/B/ST2/00920, UMO-2024/53/N/ST2/00869); Spain: Agència de Gestió d'Ajuts Universitaris i de Recerca. (AGAUR - 2023 BP 00141), Ministry of Science and Innovation (RYC2019-028510-I, RYC2020-030254-I, RYC2021-031273-I, RYC2022-038164-I), Ministerio de Ciencia, Innovación y Universidades/Agencia Estatal de Investigaci\'on (EU NextGenerationEU (PRTR-C17.I1), PID2022-142604OB-C22); Sweden: Carl Trygger Foundation (Carl Trygger Foundation CTS 22:2312), Swedish Research Council (Swedish Research Council 2023-04654, VR 2021-03651, VR 2022-03845, VR 2022-04683, VR 2023-03403, VR 2024-05451, VR 2025-05940), Knut and Alice Wallenberg Foundation (KAW 2023.0366); United Kingdom: The Binks Trust, Royal Society (NIF-R1-231091); United States of America: U.S. Department of Energy (ECA DE-AC02-76SF00515), John Templeton Foundation (John Templeton Foundation 63206), Neubauer Family Foundation.

%
%


%
%
%
\printbibliography
\clearpage

 
\begin{flushleft}
\hypersetup{urlcolor=black}
{\Large The ATLAS Collaboration}

\bigskip

\AtlasOrcid[0000-0002-6665-4934]{G.~Aad}$^\textrm{\scriptsize 102}$,
\AtlasOrcid[0000-0001-7616-1554]{E.~Aakvaag}$^\textrm{\scriptsize 17}$,
\AtlasOrcid[0000-0002-5888-2734]{B.~Abbott}$^\textrm{\scriptsize 121}$,
\AtlasOrcid[0000-0002-0287-5869]{S.~Abdelhameed}$^\textrm{\scriptsize 83b}$,
\AtlasOrcid[0000-0002-1002-1652]{K.~Abeling}$^\textrm{\scriptsize 54}$,
\AtlasOrcid[0000-0001-5763-2760]{N.J.~Abicht}$^\textrm{\scriptsize 48}$,
\AtlasOrcid[0000-0002-8496-9294]{S.H.~Abidi}$^\textrm{\scriptsize 30}$,
\AtlasOrcid[0009-0003-6578-220X]{M.~Aboelela}$^\textrm{\scriptsize 44}$,
\AtlasOrcid[0000-0002-9987-2292]{A.~Aboulhorma}$^\textrm{\scriptsize 36e}$,
\AtlasOrcid[0000-0001-5329-6640]{H.~Abramowicz}$^\textrm{\scriptsize 154}$
% ...
\bigskip
\\

% ...
$^{17}$Department for Physics and Technology, University of Bergen, Bergen; Norway.\\
% ...
$^{30}$Physics Department, Brookhaven National Laboratory, Upton NY; United States of America.\\
% ...
$^{36}$$^{(a)}$Facult\'e des Sciences Ain Chock, Universit\'e Hassan II de Casablanca;$^{(b)}$Facult\'{e} des Sciences, Universit\'{e} Ibn-Tofail, K\'{e}nitra;$^{(c)}$Facult\'e des Sciences Semlalia, Universit\'e Cadi Ayyad, LPHEA-Marrakech;$^{(d)}$LPMR, Facult\'e des Sciences, Universit\'e Mohamed Premier, Oujda;$^{(e)}$Facult\'e des sciences, Universit\'e Mohammed V, Rabat;$^{(f)}$Institute of Applied Physics, Mohammed VI Polytechnic University, Ben Guerir; Morocco.\\
% ...
$^{44}$Physics Department, Southern Methodist University, Dallas TX; United States of America.\\
% ...
$^{48}$Fakult\"{a}t Physik, Technische Universit{\"a}t Dortmund, Dortmund; Germany.\\
% ...
$^{54}$II. Physikalisches Institut, Georg-August-Universit\"{a}t G\"ottingen, G\"ottingen; Germany.\\
% ...
$^{83}$$^{(a)}$Khalifa University of Science and Technology, Abu Dhabi;$^{(b)}$New York University Abu Dhabi, Abu Dhabi;$^{(c)}$United Arab Emirates University, Al Ain;$^{(d)}$University of Sharjah, Sharjah; United Arab Emirates.\\
% ...
$^{102}$CPPM, Aix-Marseille Universit\'e, CNRS/IN2P3, Marseille; France.\\
% ...
$^{121}$Homer L. Dodge Department of Physics and Astronomy, University of Oklahoma, Norman OK; United States of America.\\
% ...
$^{154}$Raymond and Beverly Sackler School of Physics and Astronomy, Tel Aviv University, Tel Aviv; Israel.\\
% ...
$^{a}$ Also at Affiliated with an institute formerly covered by a cooperation agreement with CERN.\\
$^{b}$ Also at An-Najah National University, Nablus; Palestine.\\
$^{c}$ Also at Borough of Manhattan Community College, City University of New York, New York NY; United States of America.\\
$^{d}$ Also at Center for Interdisciplinary Research and Innovation (CIRI-AUTH), Thessaloniki; Greece.\\
$^{e}$ Also at Centre of Physics of the Universities of Minho and Porto (CF-UM-UP); Portugal.\\
$^{f}$ Also at CERN, Geneva; Switzerland.\\
% ...

\end{flushleft}


%
%

\end{document}